\documentclass[11pt,a4paper]{article}
\pdfoutput=1
\usepackage{jheppub}

\usepackage{amsmath}
\usepackage{verbatim}
\usepackage{graphicx}
\usepackage{mathrsfs}
\usepackage{appendix}
\usepackage{caption}
\usepackage{float}
\usepackage{subfig}

\newcommand{\e}{\epsilon}

\renewcommand\O{{\mathcal{O}}}
\newcommand{\be}[1]{ \begin{equation}\label{#1} }
\newcommand{\ee}{\end{equation}}
\newcommand{\bea}[1]{\begin{eqnarray}\label{#1} }
\newcommand{\eea}{\end{eqnarray}}
\newcommand{\bes}{\begin{subequations}}
\newcommand{\ees}{\end{subequations}}

\newcommand{\p}{\partial}
\renewcommand{\a}{\alpha}
\renewcommand{\b}{\beta}

\renewcommand{\th}{\theta}
\newcommand{\s}{\sigma}
\newcommand{\D}{\Delta}
\newcommand{\refb}[1]{(\ref{#1})}

\renewcommand{\>}{\rangle}

\newcommand{\non}{\nonumber}
\newcommand{\lb}{\left[}
\newcommand{\rb}{\right]}
\newcommand{\6}{^}
\newcommand{\tb}{\tilde{b}}

\title{Galilean Field Theories and Conformal Structure}

\author{Arjun Bagchi,} \author{Joydeep Chakrabortty,} \author{and Aditya Mehra.} \author{\\}

\affiliation{Indian Institute of Technology Kanpur, Kalyanpur, Kanpur 208016. INDIA. \\}

\emailAdd{abagchi, joydeep, adityams@iitk.ac.in}

\abstract{We perform a detailed analysis of Galilean field theories, starting with free theories and then interacting theories. We consider non-relativistic versions of massless scalar and Dirac field theories before we go on to review our previous construction of Galilean Electrodynamics 
and Galilean Yang-Mills theory. We show that in all these cases, the field theories exhibit non-relativistic conformal structure (in appropriate dimensions). The surprising aspect of the analysis is that the non-relativistic conformal structure exhibited by these theories, unlike relativistic conformal invariance,  becomes infinite dimensional even in spacetime dimensions greater than two. We then couple matter with Galilean gauge theories and show that there is a myriad of different sectors that arise in the non-relativistic limit from the parent relativistic theories. In every case, if the parent relativistic theory exhibited conformal invariance, we find an infinitely enhanced Galilean conformal invariance in the non-relativistic case. This leads us to suggest that infinite enhancement of symmetries in the non-relativistic limit is a generic feature of conformal field theories in any dimension.}

\preprint{}

\begin{document}

\maketitle

\newpage

\section{Introduction}

The textbook formulation of quantum field theories, especially gauge theories, is intimately linked to Lorentz invariance. Most day-to-day phenomena are, however, governed by non-relativistic physics. It is thus very useful to understand Galilean invariance and its role in quantum field theories in a deeper way than has been attempted so far in literature. There are some very startling surprises in store in the non-relativistic limit that we go on to elucidate below. It seems that there is generically an (infinite) enhancement of symmetries when one considers non-relativistic sectors of relativistic (conformal) field theories. This points to a lot of undiscovered physics and related mathematical structures in the Galilean world.   

\medskip

\noindent 
Conformal invariance and conformal field theories (CFTs) play a central role in the understanding of relativistic quantum field theories (QFTs) today. In the parameter space of all Lorentz invariant QFTs, the relativistic CFTs arise as fixed points of renormalisation group flows. QFTs thus can be understood as RG flows away from fixed points governed by conformal symmetry. The very ambitious programme of classifying (and understanding) all relativistic QFTs could thus be rephrased in terms of a classification of all relativistic CFTs. Although ambitious, this latter programme is undoubtedly more tractable than the former. Conformal bootstrap (see e.g. \cite{Simmons-Duffin:2016gjk} for a recent review of the field) has emerged as a very valuable tool in connection with the classification of relativistic CFTs. The bootstrap relies on self-consistency requirements, like conformal invariance and associativity of the operator algebra, to rule out theories. This has led to severe constraints on the parameter space of allowed CFTs.  

\medskip

\noindent 
Taking a cue from the relativistic scenario, it is tempting to attempt to understand all Galilean field theories as renormalisation group flows away from fixed points governed by the non-relativistic version of conformal invariance {\footnote{For recent progress in this direction in two dimensional BMS or GCA invariant field theories, see \cite{Bagchi:2016geg, Bagchi:2017cpu}. Some related discussion is presented in the conclusion section, at the end of the paper. Recent developments regarding a bootstrap for Warped CFTs can be found in \cite{Song:2017czq}. }}. So what is non-relativistic conformal invariance? There are several versions of non-relativistic conformal invariance and the jury is still out on which one of them is the ``true" non-relativistic conformal symmetry algebra. Below we go on to describe non-relativistic field theories and some of their conformal extensions.

\subsection*{Galilean field theories}
Galilean field theories characterised by a time direction $t$ and $(D-1)$ spatial directions $x^i$ which are spatially isotropic and homogeneous are invariant under translations of space and time and spatial rotations:
\bea{}
H: &&\quad t \to t'= t+ a, \\
P_i: && \quad x^i \to x'^i = x^i + a^i, \\ 
J_{ij}: &&\quad x^i \to x'^i = J^i_{\ j} x^j. 
\eea
These symmetries can be augmented by Galilean boosts:
\be{} 
B_i:  \quad x^i \to x'^i = x^i + v^i t
\ee
The group $\{ H, P_i, J_{ij}, B_i \}$ is called the Galilean group and can be understood as a contraction of the Poincare group. One can add a central extension to this Galilean group 
\be{}
[P_i, B_j] = \delta_{ij} M.
\ee
Here $M$ is the mass of the system under consideration. The Galilean group with the central term included is called the extended Galilean group or the Bargmann group. 

\subsection*{Non-relativistic conformal theories}
Condensed matter systems found in nature are sometimes characterised at their critical point by non-relativistic conformal field theories which exhibit a different scaling between space and time. 
\be{dil}
D: \quad t \to t'= \lambda^z t, \quad x^i \to x'^i = \lambda x^i
\ee
where $z$ is called the dynamical exponent (and $\lambda$ is a parameter of the transformation). The group $\{ H, P_i, J_{ij}, D \}$ is called the Lifshitz group and the field theories with the Lifshitz group as their underlying symmetry are called Lifshitz field theories \cite{Taylor:2015glc}. 

\medskip

The extended Galilean group together with the Galilean boosts and the dilatation operator \refb{dil} make up what is called the Schr{\"o}dinger group. For $z=2$, there is a further enhancement of this group to include a single non-relativistic special conformal generator $C$. 
\be{}
C: \quad t \to t'= \frac{t}{1+\rho  t}, \quad x^i \to x'^i = \frac{x^i}{1+\rho  t}
\ee
The Schr{\"o}dinger group derives its name from the fact that it turns out to be the maximal kinematic symmetry of the free Schr{\"o}dinger equations \cite{Hagen:1972pd, Niederer:1972zz, Henkel:1993sg}. Together with the mass, the total number of generators in this group are 13 in $D=4$. This is to be contrasted with the relativistic conformal group, which contains 15 generators in $D=4$.  

\medskip

The process of group contraction, which e.g. gets us systematically from the Poincare group to the Galilean group as we take the speed of light to infinity, preserves the number of generators. It is clear that the Schr{\"o}dinger group would not appear as a contraction of the 
relativistic conformal group, as the number of generators of the two groups are not the same. The contraction of the relativistic conformal algebra will lead us to what is called the finite Galilean Conformal Algebra (GCA) \cite{Bagchi:2009my}. This symmetry, governed by the GCA is thus in a sense the ``true" non-relativistic limit of the relativistic conformal symmetry. We will be concerned with this version of non-relativistic conformal symmetry in our work. A detailed review of the GCA would appear in Sec~2. 

\subsection*{Motivations of this paper}
In this paper, we will investigate Galilean field theories and their emergent conformal structure. We will consider relativistic field theories that enjoy conformal invariance and follow a systematic non-relativistic limit in order to define the Galilean field theories. We will then investigate whether the conformal structure carries over to the non-relativistic case. The symmetry algebra of interest in these non-relativistic theories would be the GCA. This paper is a continuation and elaboration of the earlier works \cite{Bagchi:2014ysa, Bagchi:2015qcw}, where we have addressed the Galilean Electrodynamics and Galilean Yang-Mills theories. Specifically, we would first attempt to understand the non-relativistic versions of scalar fields and fermionic fields. These matter fields would then be added to the gauge fields as we build towards a complete understanding of interacting field theories in the Galilean regime. 

We would be studying of symmetries of the equations of motion (EOM) of these individual systems, as opposed to looking for the symmetries of the respective actions. This is because defining actions are often problematic in the non-relativistic case. One of the crucial reasons behind this is the fact that the metric degenerates in the non-relativistic limit and Riemannian structures associated with the spacetime geometry need to be replaced by Newton-Cartan structures. We will not have much to say about actions in this paper. In order to escape the complications posed for defining actions, we shall restrict ourselves to investigating the limits on relativistic EOM and see how to reduce to the Galilean EOM. Using methods outlined in earlier work \cite{Bagchi:2014ysa, Bagchi:2015qcw}, we shall then investigate the symmetries of these equations. An equivalent treatment of our investigations in this paper should be possible in an action formulation. We leave this to future work. 

The finite GCA, as we will go on to review in the next section, admits infinite dimensional extensions in all spacetime dimension (which we will at times refer to as the infinite GCA). In dimensions $D>2$, this is an enhancement of symmetries that is found in the non-relativistic limit and  is {\it{not}} the result of a contraction of the relativistic conformal algebra, which is $\textit{SO(D,2)}$. All the non-relativistic systems that we consider in this paper are descendants of parent relativistic conformal field theories. We find that all of these Galilean field theories are not only invariant under the finite GCA, which is the contraction of  symmetries of the parent system, they are invariant under the full infinite dimensional extension of the GCA {\footnote{Whether or not there exist non-relativistic conformal field theories which do not have a relativistic origin is an important question that we don't address in this current paper. See the concluding section for some remarks in this direction.}}.        

The above is a tantalising hint that there is a {\em{generic}} enhancement of symmetries in the non-relativistic limit of any conformal field theory and that this enhancement is {\em{infinite dimensional}}. Infinite symmetries are a pointer to the integrability \cite{Beisert:2010jr} and ultimately the complete solvability of a system. So it seems that the non-relativistic limit of any conformal field theory would lead to a Galilean sub-sector that is integrable. This may have profound consequences and can perhaps be useful to understand better the theory of strong interactions \cite{Brambilla:2004jw, Soto:2006zs}. 

\subsection*{Outline of this paper}

With the above motivations in mind, we would go on to address the construction of Galilean field theories in this paper. The next section contains a comprehensive review of Galilean conformal symmetry and its representation theory. In Sec 3, we address free fields. We start out with the simplest of examples of the scalar field. Fermions are then addressed. All of this is new material. We then go on to review our earlier construction of Galilean Electrodynamics \cite{Bagchi:2014ysa}, which built on seminal work in 1970's by Le Bellac and Levy-Leblond \cite{LBLL}. Work on non-relativistic field theories, similar in spirit are \cite{Duval:2014uoa, Festuccia:2016caf, Bergshoeff:2015sic, Bleeken:2015ykr}. In Sec 4, we address how to couple different fields together and build interacting Galilean theories. In this section, we consider scalar and then fermionic fields coupled to Galilean Electrodynamics and study the symmetries of the EOM in each case.  In Sec 5, we continue to build on interacting theories now by considering Galilean Yang-Mills theory. After a brief recap of the construction of \cite{Bagchi:2015qcw}, we focus on the Galilean $\textit{SU(2)}$ theory with fermions. Section 6 contains a summary of our paper along with further remarks. We devote three appendices to a detailed discussion of generalised limits in the Galilean theory.  

\section{Galilean conformal field theories: General considerations}
In this section we will provide a quick summary of the GCA along with a description of its representation theory. The details of the representation theory constructed here will be used in the coming sections to find the symmetries of the EOM of several non-relativistic field theories.

\subsection{Algebraic aspects}
In order to obtain the GCA in any arbitrary spacetime dimensions, we perform an In{\"o}n{\"u}-Wigner contraction on the conformal algebra. As stated in the introduction, this is the systematic procedure which generates the Galilean algebra from the Poincare algebra. We will perform the limit at the level of spacetime. Lorentz covariance breaks down in the process of going to the Galilean framework which entails taking the speed of light to infinity. We will parametrise this by: 
\be{sptmcontract}
x_i \rightarrow \epsilon x_i, \, \, t \rightarrow t , \, \, \epsilon \rightarrow 0.
\ee
This process e.g. on the Lorentz boost yields:  
$$J_{0i} = t \p_i + x_i \p _t \mapsto \epsilon ^{-1} t \p _i + \epsilon x_i \p _t.$$  
In order to extract the Galilean boost, we will define
$$B_{i} = \lim_{\e\to 0} \e J_{0i} = t \p_i,$$ 
making the generator finite in the limit. The algebra thus obtained in the limit from the relativistic conformal algebra is called the finite GCA. The generators of this algebra can be written as:
\begin{subequations}\label{suggen}
\bea{}
&& L^{(n)} = -t^{n+1} \p_t - (n+1) t^n x^i \p_i,  \quad M^{(n)}_i = t^{n+1} \p_i ~~ \mbox{for}~~ n=0, \pm 1 \\
&&  J_{ij} = -x_{[i} \p_{j]}.
\eea
\end{subequations}
A more familiar identification is $L^{(-1,0,1)}=H,D,K$ and $M^{(-1,0,1)}_i =P_{i},B_{i},K_{i}$ where $H, D$ and $K$ are respectively the Galilean Hamiltonian, dilatation and temporal special conformal transformation. On the other hand, $P_i, B_i$ and $K_i$ represent momentum, Galilean boost and spatial special conformal transformation. $J_{ij}$ generates homogeneous $SO(D-1)$ rotations. The full algebra can be written as      
\bea{GCA}
&& [L^{(n)}, L^{(m)}] = (n-m) L^{(n+m)}, \quad [L^{(n)}, M^{(m)}_i] = (n-m) M_i^{(n+m)},\non\\
&& [M^{(n)}_i, M^{(m)}_j]=0, \quad [L^{(n)}, J_{ij}] = 0, \quad [J_{ij},M^{(n)}_k] = M^{(n)}_{[j} \delta_{i]k}
\eea
with $n,m=0,\pm 1$. The algebra admits the usual Virasoro central charge $\frac{c}{12} (n^3-n) \delta_{n+m,0}$ in the $[L,L]$ commutator. 

One of the most interesting observations \cite{Bagchi:2009my} about this algebra is that even if we let the index $n$ of \eqref{suggen} run over all integers, the algebra  \eqref{GCA} closes. This is thus an infinite extension of the algebra which was previously obtained by a contraction from the finite relativistic conformal algebra  {\footnote{It is worth pointing out here that there is an even larger extension \cite{Bagchi:2009my} where the rotation generator is given an infinite lift according to $$J_{ij}^{(n)} = - t^n x_{[i} \p_{j]}.$$ However, we will not be interested in this extension as the systems we go on to examine do not exhibit this symmetry. It is unclear to us presently why this symmetry is absent.}}. We will refer to this infinite dimensional algebra as the GCA from now on. The GCA is thus infinite dimensional for all spacetime dimensions, unlike the relativistic conformal algebra which is infinite only in $D=2$. One of the novel claims of \cite{Bagchi:2009my} was that there would be an infinite enhancement of symmetries in the non-relativistic limit of all relativistic conformal field theories. In the later sections of this paper, we provide evidence to this conjecture.

\subsection{Representation theory} \label{repth}
We will construct the representations of GCA in a method similar to that of the relativistic conformal algebra.  The states in the theory are labelling by their weights under the dilatation operator $L^{(0)}$ \cite{Bagchi:2009ca, Bagchi:2009pe}:
\bea{}
L^{(0)} | \Phi \rangle = \Delta | \Phi \rangle.
\eea
 The operators $L^{(n)}, M_i^{(n)}$ lower the weights for positive $n$  while they raise the weights for negative $n$: 
\bea{}
L^{(0)} L^{(n)} | \Phi \rangle = (\Delta -n) L^{(n)} | \Phi \rangle, \quad L^{(0)} M_i^{(n)} | \Phi \rangle = (\Delta -n) M_i^{(n)} | \Phi \rangle.
\eea
If we wish to have a bounded spectrum of $\Delta$, we cannot lower the weights indefinitely. Hence we define the notion of GCA primary states as 
\bea{}
L^{(n)} | \Phi \rangle_p = M_i^{(n)} | \Phi \rangle_p = 0 \quad \forall n >0.
\eea
We now have two different options in labelling these states further. Since 
\bea{}
[L^{(0)}, M_i^{(0)}] = 0, \quad [L^{(0)}, J_{ij}] = 0, \quad [M_k^{(0)}, J_{ij}] \neq 0. 
\eea
we see, the states can be labelled either by $M_{i}^{0}$ or $J_{ij}$ but not with both of them. In this paper, we would label our states with $J_{ij}$ as we would be interested in operators with spin. These GCA primaries would be called scale-spin primaries. The following discussion closely follows earlier work \cite{Bagchi:2014ysa, Bagchi:2015qcw}. For representations labeled by the boost generator, the reader is referred to \cite{Bagchi:2009ca} (and to \cite{Bagchi:2009pe} for the $D=2$ case). 

Now returning to the above described scale-spin representations, the action of $J_{ij}$ on the primaries in a particular representation is given as
\bea{}
J_{ij} | \Phi \>_p = \Sigma_{ij} | \Phi \rangle_p.
\eea
The representations of the GCA are built by acting with the raising operators ($L^{(n)}, M_i^{(n)}$ for negative $n$) on these primary states. 
We postulate the existence of a state-operator correspondence in the GCA in close analogy with relativistic CFTs{\footnote{This is not a strict requirement, but it makes life substantially simpler if we can move between states and operators in our analysis. The state-operator map can be motivated as arising from the limit from the parent relativistic CFT.}}: 
\be{}
| \Phi \rangle_p = \Phi (0,0) | 0 \rangle.
\ee
We will move interchangeably between states and operators in the discussion that follows. The action of the finite part of GCA on the above described primaries is given as: 
\bes 
\bea{}
&&\left[L^{(-1)}, \Phi(t,x)\right] = \partial _t \Phi(t,x) \label{ham1},~ 
\left[M^{(-1)}_i, \Phi(t,x)\right]= -\partial _i \Phi(t,x) \label{mom1},\\
&&\left[J_{ij}, \Phi(0,0)\right] =  \Sigma _{ij} \Phi(0,0),~
\left[L^{(0)}, \Phi(0,0)\right] = \Delta \Phi(0,0), \\
&&\left[L^{(+1)}, \Phi(0,0)\right] = 0 = \left[M^{(+1)}_i, \Phi(0,0)\right].
\eea
\ees
The actions of $J_{ij}, L^{(0,1)}$ and $M^{(1)}_i$ can be worked out on an operator at general spacetime points $(t,x)$ as follows. According to our conventions,
\bea{}
\Phi(t,x) = U \Phi(0,0) U^{-1},
\eea
where $U= e^{t L^{(-1)} - x^i  M^{(-1)}_i}$. For a general GCA element $ \O$, we therefore use:
\bea{}
\left[ \O, \Phi(t,x)\right] = U \left[ U^{-1} \O U , \Phi(0,0)\right] U^{-1} \nonumber
\eea
and then exploit the Baker-Campbell-Hausdorff formula and the commutation relations of the GCA to evaluate $U^{-1} \O \, U$. This is straightforward for $J^{ij}$ and $L_{(0)}$:
\bea{} \label{jacspin11}
&& \left[J_{ij}, \Phi(t,x)\right] =  x_{[i} \p_{j]} \Phi(t,x)+ \Sigma_{ij} \Phi(t,x), \non \\
&& \left[L^{(0)}, \Phi(t,x)\right]= (t \partial _t +x^i \partial_i + \Delta) \Phi (t,x).
\eea
An important point is the role of boost on the operators. For that, we look at the  relevant Jacobi identity:
\bea{} 
\left[J_{ij}, \left[ B_k , \Phi (0,0)\right] \right]=  \lb B_k, \Sigma_{ij} \Phi (0,0) \rb + \delta_{k [i} \lb B_{j]}, \Phi (0,0) \rb.\eea
In order to solve the above equation for fields having different spins we use:
\be{bost}
  [B_{k},\Phi(0,0)]= a\varphi +b\sigma_{k}\chi +\tb \sigma_{k}\phi + s A_{k}+rA_{t}\delta_{ki} + \ldots,
\ee
where $\Phi=\{\varphi,\phi, \chi, A_{t}, A_{i}\}$ and $\varphi$ is a scalar field, $(\chi,\phi)$ are 2-component fermions and $A_i, A_t$ are spatial and temporal components of a vector field. The dots indicate higher spin fields. For example, if we take $\Phi(0,0) = A_{t}$, then to satisfy the Jacobi identity we have to take $ [B_{k},\Phi(0,0)]=s A_{k}$. The action of boost on other fields can be shown in a similar manner. We are yet to determine the constants appearing in (\ref{bost}). The determination of these constants demands inputs from dynamics. We will clarify how these constants get fixed for particular physical systems.
\\
The action of boost on general operator $\Phi(x,t)$ at finite spacetime points gives:
\bea{}\label{boq} [B_{k},\Phi(t,x)] = -t\p_{k}\Phi(t,x)+ U[B_{k},\Phi(0,0)]U^{-1}.
\eea
The action of $L^n$ and $M_{i}^{n}$ on a general operator $\Phi(t,x)$ of
weight $\Delta$ at arbitrary spacetime points is thus given by 
\bea{inftr}   
&& [L^{(n)},\Phi(t,x)]= \left(t^{n+1}\p_{t}+(n+1)t^{n}(x^{l}\p_{l} +\Delta)\right)\Phi(t,x) -t^{n-1}n(n+1)x^{k} U[M_{k}^{(0)},\Phi(0,0)]U^{-1},\non \\
&& [M_{l}^{(n)},\Phi(t,x)]= -t^{n+1}\p_{l}\Phi(t,x)+(n+1)t^{n}U[M_{l}^{(0)},\Phi(0,0)]U^{-1}.
\eea  
The action of $L^{(n)}$ and $M_{i}^{(n)}$ on $\Phi(t,x)$ will be used to find the invariance of EOM of a particular theory. 

\section{Free Galilean field theories}
\subsection{Scalar fields}
As we just reviewed, the non-relativistic limit of conformal algebra can be systematically constructed by using the scaling on spacetime (\ref{sptmcontract}) and it has been claimed that there is an infinite enhancement of symmetries in any dimensions in this Galilean limit. It has been shown in  recent work that some systems, viz. Galilean Electrodynamics \cite{Bagchi:2014ysa} and Galilean Yang-Mills theories \cite{Bagchi:2015qcw}, exhibit the aforementioned infinite dimensional symmetry. In this paper, we aim to build upon this and show that any relativistic conformal field theory exhibits this phenomenon. To this end, we first look at free Galilean scalars and fermions. We will then use this construction to build interacting theories in the Galilean regime, and especially to couple matter to Galilean gauge theories that have been earlier discussed in \cite{Bagchi:2014ysa, Bagchi:2015qcw}. 
We will only consider massless theories as the presence of mass breaks conformal invariance. For the purposes of this paper, we shall confine our attention to $D=4$ unless otherwise stated. 

\subsubsection*{Relativistic Scalar theory}
We start with a brief description on relativistic conformal transformations and the conformal invariance of massless scalar field theory. In order to set notation, we recall the conformal algebra in $D=4$ has the following generators: 
\begin{flalign}
\mbox{Poincare generators:} \quad \tilde P_i= \p_i, \tilde H=-\p_t, \tilde J_{i j} = x_{[i} \p_{j]}, \tilde B_i = x_i \p_t + t \p_i \\
\mbox{Conformal generators:} \quad \tilde D = -x \cdot \p, \ \tilde K_\mu =  - (2 x_\mu (x \cdot \p)  - (x \cdot x) \p_\mu) \label{DK}
\end{flalign}
The conformal algebra in $D$ dimensions is isomorphic to $\mathfrak{so}(D,2)$. To highlight the differences with the GCA \refb{GCA}, few important commutation relations are indicated below: 
\be{rel-al}
[\tilde P_i, \tilde B_j]= - \delta_{ij} \tilde H, \, \, [\tilde B_i, \tilde B_j] =  \tilde J_{ij}, \, \, [\tilde K_i, \tilde B_j] = \delta_{ij} \tilde K, \, \, [\tilde K_i, \tilde P_j] = 2 \tilde J_{ij} + 2 \delta_{ij} \tilde D. 
\ee
The right hand side of all these commutators are zero in the finite GCA, while all other commutators stay the same. 

\medskip

We now describe the conformal transformations of fields and review the formalism of checking for conformal symmetry from the point of view of the EOM following \cite{Jackiw:2011vz}. Poincare transformations of a multi-component field $\Phi$:
\be{poin}
{\delta}_\mu^{\mbox{\tiny{P}}} \, \Phi (x) = \p_\mu \Phi (x), \quad {\delta}_{\mu \nu}^{\mbox{\tiny{L}}} \, \Phi (x) = (x_\mu \p_\nu - x_\nu \p_\mu + \Sigma_{\mu\nu}) \Phi(x).
\ee
Transformations under scaling takes the following form:
\be{st}
{\delta}^{\mbox{\tiny{D}}}  \, \Phi (x) = (x \cdot \p + {\tilde{\Delta}}) \Phi,
\ee
where ${\tilde{\Delta}}$ is the scaling dimension of the field $\Phi$. In $D$ dimensions, the scaling dimension of a scalar field and a vector field are ${\tilde{\Delta}}=\frac{D-2}{2}$, whereas for a fermionic field ${\tilde{\Delta}}=\frac{D-1}{2}$. The transformation under special conformal transformation is given as:
\be{sctt}
{\delta}^{\mbox{\tiny{K}}}_\mu  \, \Phi (x) = ( 2 x_\mu (x \cdot \p) - x^2 \p_\mu + 2 {{\tilde{\Delta}}} x_\mu - 2 x^\nu \Sigma_{\nu \mu} )  \, \Phi (x).
\ee
Consider now a real massless scalar field $\varphi(x)$ in $D$-dimensional spacetime. The Lagrangian is given as
\be{} \mathcal{L} = -\frac{1}{2}\p_{\mu}\varphi(x)\p^{\mu}\varphi(x),\ee
The corresponding equation of motion is:
\be{sceom} \p^{\mu}\p_{\mu}\varphi(x)=0.\ee
The above Lagrangian and equation of motion are manifestly invariant under Poincare transformations. We wish to examine the action of the dilatation and special conformal transformation on the equation of motion:
\be{}\delta_{D}(\p^{\mu}\p_{\mu}\varphi(x))=2\p\6\tau\p_{\tau}\varphi=0,~~\delta^{\sigma}_{K}(\p^{\mu}\p_{\mu} \varphi(x))=4\left({\tilde{\Delta}} -\frac{D-2}{2}\right)\p^{\sigma}\varphi.\ee
We have already stated that ${\tilde{\Delta}}=\frac{D-2}{2}$ for scalar fields. This indicates that massless scalar field theory is scale invariant as well as conformal invariant in all dimensions $D$.

\subsubsection*{Galilean Scalars}
Now we work out the details of the non-relativistic limit of the free massless scalar field theory we discussed above. This will provide the prototype for all the sections to follow, where we will examine the symmetries of EOM arising from different field theories. The non-relativistic EOM for the free scalar theory can be found by using the scaling of the co-ordinates (\ref{sptmcontract}) on the relativistic equation of motion \refb{sceom}. This simply gives rise to:
\be{nrs} \p^{i}\p_{i}\varphi(x)=0.\ee 
\refb{nrs} is obviously a rather uninteresting example as all the time derivatives have disappeared from the equation. But this would be instructive as an example for what is to follow. 
For checking the invariance of (\ref{nrs}), we will consider the transformation of the field variable $\Phi(t, x)$ under an infinitesimal transformation generated by $Q$. We will check whether the transformed field variable satisfies the EOM. The infinitesimal change in the field variable is given by: 
\bea{}
\delta_{\varepsilon} \Phi(t,x) = \lb  \varepsilon Q , \Phi (t,x) \rb,
\eea
where $\varepsilon$ is the parameter corresponding to particular transformation under consideration. If $Q$ generates a symmetry of the EOM, we would have:
\bea{prosc}
\square \delta \Phi(t,x) =\square \lb  Q , \Phi (t,x) \rb =0,
\eea
where $\square \Phi(x,t)=0$ is the schematic form for EOM. For the present case, $\Phi(t,x)$ is a scalar field $\varphi(t,x)$.  We will check for the  symmetries of the EOM. For checking the invariance under $K,D$ and $K_{l}$, we use (\ref{inftr}):
\bea{} \p\6i \p_i [K,\varphi]=\p\6i \p_i [t^{2}\p_{t}\varphi +2tx^{l}\p_{l}\varphi +2t\Delta\varphi] =4t\p^{i}\p_i \varphi +2tx^{l}\p_{l} (\p\6i \p_i \varphi)=0,\\
 \p\6i \p_i [D,\varphi] =\p\6i \p_i [t\p_{t}\varphi +x^{l}\p_{l}\varphi +\Delta\varphi] = 2\p^{i}\p_i \varphi +x^{l}\p_{l} (\p\6i \p_i \varphi)=0,~~~~~~~~~~\\ \p\6i \p_i [K_{l},\varphi]=-t^{2}\p_{l}(\p\6i \p_i\varphi)=0.~~~~~~~~~~~~~~~~~~~~~~~~~~~~~~~~~~~~~~~~~~~~~~~~~~~~~~~\eea
As expected, the non-relativistic equation of motion of the massless scalar theory is invariant under the finite GCA. Note here that the differential operator form of the dilatation operator $D$ \refb{suggen} does not change from the relativistic one $\tilde{D}$ \refb{DK}. Hence the dilatation eigenvalue ${\tilde{\Delta}}$ does not change as we go from the relativistic theory to the non-relativistic sector, where we have labeled it as $\Delta$. From now on, we shall not distinguish between the two, i.e. we set ${\tilde{\Delta}} = \D$. 

The non-trivial part of the analysis of symmetries for the equation \refb{nrs} is the proof that it exhibits an infinite dimensional symmetry under the extended GCA. By using the details of the representation theory discussed earlier, specifically \refb{inftr}, we check the transformation of (\ref{nrs}) under the generators of the infinite algebra $L^{(n)}$ and $M^{(n)}_{i}$: 
\bea{}  \non \p\6i \p_i [L^{(n)},\varphi] =\p\6i \p_i [t^{n+1}\p_{t}\varphi +(n+1)t^{n}x^{l}\p_{l}\varphi +(n+1)t^{n}\Delta\varphi] \\=(n+1)t^n [2\p^{i}\p_i \varphi +x^{l}\p_{l} (\p\6i \p_i \varphi)]=0,~~~~~~~~~~~~~\\ 
\p\6i \p_i [M_{l}^{(n)},\varphi]= -t^{n+1}\p_{l}(\p\6i \p_i\varphi) =0.~~~~~~~~~~~~~~~~~~~~~~~~~~~~~~~~
\eea
The EOM are hence invariant under infinite GCA in all dimensions. \\

\subsection{Fermionic fields}
Next we consider massless fermions. The massless relativistic Dirac Lagrangian is  
\bea{}  \mathcal{L}= i\bar{\Psi}(x)\gamma^{\mu}\p_{\mu}\Psi(x), \eea
where $\Psi(x)$ is a 4-component Dirac fermion. The equation of motion is:
\bea{feom} i\gamma^{\mu}\partial_{\mu}\Psi(x)=0,\eea
where $\gamma^\mu$ are the gamma matrices which follow the Clifford algebra: $\lbrace \gamma^{\mu},\gamma^{\nu}\rbrace =-2\eta^{\mu\nu}.$The theory is, by construction, Poincare invariant (\ref{poin}). We will focus on dimensions $D=4$ in this analysis as we are going to use particular representations of the gamma matrices. The action of scale (\ref{st}) and special conformal transformations (\ref{sctt}) on \refb{feom} are:  
\be{} i\gamma^{\mu}\partial_{\mu}\delta_{D}\Psi(x)=0,~~ i\gamma^{\mu}\partial_{\mu}\delta^{\sigma}_{K}\Psi(x)=
 2i\gamma^{\sigma}(\Delta -\frac{3}{2})\Psi = 0, \ee
where we have used  $\gamma^{\mu}\Sigma^{\sigma\tau}=(\eta^{\mu\sigma}\gamma^{\tau}-\eta^{\mu\tau}\gamma^{\sigma}+
\Sigma^{\sigma\tau}\gamma^{\mu})$ and  $\gamma^{\mu}\gamma_{\mu}=-4$ in the intermediate steps.  
\refb{feom} is thus conformally invariant. 

\subsubsection*{Galilean Fermions} \label{gdft}
We now look into the non-relativistic limit of massless Dirac field theory and study its symmetry structure. We will decompose the  fermionic field $\Psi(t,x)$ into two component spinors: 
\be{compbr} \Psi = \begin{pmatrix} \phi \\
    \chi \\
\end{pmatrix} \ee 
and then take the non-relativistic limit. We will perform our analysis in Pauli-Dirac representations{\footnote{We consider fermions in the Weyl representation in Appendix \ref{ApB} and show that the non-relativistic limit there leads to the same results.}}:
\be{gama1} 
\gamma^{0}= 
\begin{pmatrix} 1 & 0 \\
    0 & -1 \\
\end{pmatrix}
,~~
\gamma^{i}= 
\begin{pmatrix} 0 & \sigma^{i} \\
    -\sigma^{i} & 0 \\
\end{pmatrix}, \ee
where $\sigma^{i}$ are the Pauli matrices. The relativistic EOM for the spinors $(\phi,\chi)$ are given by
\be{mde} i\p_{t}\phi + i\sigma.\p\chi =0,~
 i\sigma.\p\phi +i\p_{t}\chi =0.\ee
To get the non-relativistic equation, we will scale the spinors as 
\be{nrsc}\phi\rightarrow \phi,~ \chi\rightarrow\epsilon\chi,\ee 
along with scaling of spacetime (\ref{sptmcontract}).
We would have also taken $\phi\rightarrow \e\phi,~ \chi\rightarrow\chi$ as our choice of scaling of the spinors. But (\ref{mde}) has a $\mathbb{Z}_2$ symmetry which means that the exchange of two spinor will have no effect on the EOM. 

The action of Galilean boosts of fermions can be obtained by taking the non-relativistic limit on the Lorentz boosts in (\ref{poin}): 
\be{}\delta^{B}_{0i}\phi=-t\p_{i}\phi,~~\delta^{B}_{0i}\chi=-t\p_{i}\chi+\frac{\sigma^{i}}{2}\phi.\ee
Comparing with (\ref{bost}), the arbitrary constants for this theory that label the scale-spin representations are fixed to $b = 0,~ \tb= \frac{1}{2}$. This, along with the assumption that the dilatation eigenvalue $\Delta$ is inherited unchanged from the relativistic theory as argued before, are the required input from dynamics in the general Galilean conformal representation theory elaborated in Sec 2. 
In the non-relativistic limit, massless Dirac equation reduces to
\bea{eomnr} i\sigma.\p\phi =0 , ~i\p_{t}\phi +i\sigma.\p\chi =0. \eea
The relativistic EOM \refb{feom} were invariant under the relativistic conformal group in $D = 4$. A scaling limit of these equations leads to (\ref{eomnr}) and the same limit on the conformal group leads to the finite GCA. It is thus expected that the non-relativistic equations will be invariant under the finite GCA. We will now check this from an intrinsic Galilean perspective with required input from dynamics. At an operational level, this means that  we check the symmetries of the EOM using (\ref{prosc}) along with (\ref{inftr}). The EOM under scale transformation remains trivially invariant.  Checking for the invariance under spatial and temporal special conformal transformations gives 
\bea{ink0} i\sigma\6i\p_{i}[K_{l},\phi(t,x)]= 2ibt\sigma\6i\sigma_l\p_{i}\chi,~
 i\sigma\6i\p_{i}[K,\phi(t,x)]= -2ib\sigma^{i}\p_i(x^l \sigma_{l}\chi).\eea
Similarly, we get
\bes
\bea{} 
&& i\p_{t}[K_{l},\phi]+i\sigma^{i}\p_{i}[K_{l},\chi]=-2ti(1-2\tb)\p_{l}\phi +2ib\p_{t}(t\sigma_{l}\chi),\label{ink1}\\
&& i\p_{t}[K,\phi]+i\sigma^{i}\p_{i}[K,\chi]=2ix^{l}(1-2\tb)\p_{l}\phi + 2i(\Delta-\tb\sigma^{i}\sigma_{i})\phi -2bx^{l}\sigma_{l}(i\p_{t}\chi)\label{ink2} \eea
\ees
Now $\{b,\tb\} =\{0, \frac{1}{2}\}$ from above, and $\Delta=\frac{3}{2}$. Substituting this in (\ref{ink0}-\ref{ink2}), we get:
\bea{} 
&&i\sigma\6i\p_{i}[K_{l},\phi(t,x)]= 0,~
 i\sigma\6i\p_{i}[K,\phi(t,x)]= 0,\\
&& i\p_{t}[K_{l},\phi]+i\sigma^{i}\p_{i}[K_{l},\chi]=0, \\
&& i\p_{t}[K,\phi]+i\sigma^{i}\p_{i}[K,\chi]=2i(\Delta-\frac{1}{2}\sigma^{i}\sigma_{i})\phi=2i(\Delta-\frac{3}{2})\phi=0.\eea
Therefore, the equations (\ref{eomnr}) are invariant under the finite GCA, as was expected. 

\medskip

\noindent Now we extend the analysis to the infinite modes of the GCA. Under $M_{l}^{(n)}$, we have 
\bea{} i\sigma\6i \p_{i}[M_{l}^{(n)},\phi]= 0,~ i\p_{t}[M_{l}^{(n)},\phi]+i\sigma^{i}\p_{i}[M_{l}^{(n)},\chi]=0.\eea
The EOM are invariant under all $M_{l}^{(n)}$. Similarly, checking for the invariance under $L^{(n)}$:
\be{}  i\sigma\6i\p_{i}[L^{(n)},\phi]= 0,~
i\p_{t}[L^{(n)},\phi]+i\sigma^{i}\p_{i}[L^{(n)},\chi]=in(n+1)t^{n-1}(\Delta-\frac{3}{2})\phi=0.\ee
Hence,  the EOM are also invariant under all $L^{n}$.

\subsection{Vector fields}\label{vfr}
The study of Galilean electrodynamics stemmed from the work of Le Bellac and Levy-Leblond in \cite{LBLL}. Our analysis in \cite{Bagchi:2014ysa} was motivated by the search of symmetry in this non-relativistic version of electrodynamics. Maxwell equations in relativistic electrodynamics are classically conformally invariant in $D=4$. It is thus natural to expect that the non-relativistic version of Maxwell's theory would exhibit non-relativistic conformal invariance in $D=4$. The emergence of the finite GCA as symmetries is almost guaranteed if one performs the limit properly. The non-triviality is, as always, the existence of infinitely extended symmetries in the non-relativistic limit. 

In \citep{Bagchi:2014ysa}, our discussion of Galilean electrodynamics was based on the potential formulation. In addition to the scaling of spacetime, here we needed to prescribe how the scalar and vector potential scaled in the Galilean regime. In keeping with the earlier analysis of Le Ballac and Levi-Leblond, we found that we could scale the 4-vector potential $A_{\mu}$ in two different ways: 
\be{sca} A_t \rightarrow A_t, \, \, A_i \rightarrow \epsilon A_i ~~\text{and}~~ A_t \rightarrow \epsilon A_t, \,  \, A_i \rightarrow  A_i.\ee 
The first limit is the Electric limit, where the electric effects are much stronger than the magnetic effects ($ |\mathbf{E}| \gg | \mathbf{B}|$). The second limit is known as Magnetic limit where the situation is reversed ($ |\mathbf{E}| \ll | \mathbf{B}|$). From the point of view of the underlying spacetime, the origin of the two different limit can be traced to the degeneration of the spacetime metric. In the Galilean world, contravariant and covariant vectors behave very differently since there is no spacetime metric to take one description to the other. The above scalings appear because we treat the original relativistic four potential once as a contravariant vector and then as a covariant vector before performing the limit.  

The EOM of Galilean electrodynamics (in absence of sources) in these two sectors are:
\bes
\bea{}
\hspace{-2cm}\mbox{{\bf{Electric limit}}:}\qquad &&\partial^i \partial_i A_t = 0, \quad \partial ^j \partial_j A_i - \partial _i \partial_j A^j + \partial_t \partial _i A_t= 0; \label{Eeom} \\
\mbox{{\bf{Magnetic limit}}:} \qquad &&(\partial ^j \partial_j)A_{i}-\partial_{i}\partial_{j}A^{j} =0, \quad (\partial^i \partial_i)A_{t}-\partial_{i}\partial_{t}A^{i}= 0. \label{Meom}
\eea \ees
For the invariance of the EOM under the whole GCA, we took the required inputs from the representation theory (\ref{inftr}) and used the schematic form (\ref{prosc}). We found that this theory has the underlying infinite dimensional symmetry of the GCA in $D = 4$. For further details, we leave it to the readers to look at \cite{Bagchi:2014ysa}. For other recent works in this direction, the reader is pointed to \cite{Duval:2014uoa, Bergshoeff:2015sic, Festuccia:2016caf}. The conclusions section contains a discussion of how the analysis of \cite{Festuccia:2016caf} differs from our earlier work \cite{Bagchi:2014ysa}.  
\section{Interacting Galilean field theories I}
\subsection{Yukawa theory}
We have so far dealt exclusively with free theories. Sceptic may argue that the infinite symmetry of the non-relativistic massless free scalars and fermions theories as well as Galilean electrodynamics arose precisely because of the absence of interactions. To alleviate such concerns, we begin our construction of interacting Galilean theories. In this section, we will construct non-relativistic versions of Yukawa theory, scalar electrodynamics and electrodynamics with fermions. 

We will begin our discussion with the action of massless relativistic Yukawa theory
\be{aY} S =\int d^{4}x~ [-\frac{1}{2}\p^{\mu}\varphi\p_{\mu}\varphi +i \bar{\Psi}\gamma^{\mu}\p_{\mu}\Psi -g\bar{\Psi}\varphi\Psi].  \ee   
Here, the underlying symmetry is $U(1)$ global and $\varphi$ and $\Psi$ are the singlet scalar, and  non-singlet Dirac fields respectively with  coupling constant $g$. The EOM are
\be{eomyt} 
\p^{\mu}\p_{\mu}\varphi -g\bar{\Psi}\Psi =0,\qquad
i\gamma^{\mu}\p_{\mu}\Psi -g\varphi\Psi=0.
\ee
The action of dilatation (\ref{st}) on EOM is given as  
\bes
\bea{sti}
 \delta_{D}(\p^{\mu}\p_{\mu}\varphi-g\bar{\Psi}\Psi) = (\Delta_1 -1)\p^{\mu}\p_{\mu} \varphi - 2(\Delta_2 -\frac{3}{2})g\bar{\Psi}\Psi=0,\\
\delta_D(i\gamma^{\mu}\p_{\mu}\Psi -g\varphi\Psi)= -(\Delta_1 -1)g\varphi \Psi=0.
\eea
\ees
Thus, in $D=4$, the theory is scale invariant only if we take the scaling dimensions $\Delta_1 =1$ for the scalar field and $\Delta_2 =\frac{3}{2}$ for the Dirac field. This can be seen both from the EOM and the action. The action of special conformal transformation (\ref{sctt}) on EOM is 
\bes
\bea{} \delta_K^{\sigma}( \p^{\mu}\p_{\mu}\varphi -g \bar{\psi}\psi) = (2\Delta_1 -2)[x^{\sigma}\p^{\mu}\p_{\mu}\varphi +2\p^{\sigma}\varphi]- (4\Delta_2 -6)x^{\sigma}g\bar{\Psi}\Psi =0, 
\\
\delta_K^{\sigma}(i\gamma^{\mu}\p_{\mu}\Psi -g\varphi\Psi)= (2\Delta_2 -3)i\gamma^{\sigma}\Psi -2(\Delta_1 -1)gx^{\sigma}\varphi\Psi=0.
\eea
\ees
With the input of the scaling dimensions of the fields, we see that the EOM (\ref{eomyt}) are also invariant under special conformal transformation. 
\subsection*{Galilean Yukawa theory}
In constructing the non-relativistic limit of Yukawa theory, we will follow the analysis in Sec[\ref{gdft}], where we decomposed the Dirac spinor into two component spinors and used Pauli-Dirac representation of gamma matrices. The relativistic EOM (\ref{eomyt}) can be written down  in terms of $(\phi,\chi)$:
\be{eomyt1} \p^{\mu}\p_{\mu}\varphi -g(\phi^{\dagger}\phi -\chi^{\dagger}\chi)=0,~
i\p_{t}\phi +i\sigma^{i}\p_{i}\chi -g\varphi\phi=0,~
i\p_{t}\chi +i\sigma^{i}\p_{i}\phi +g\varphi\chi=0.\ee
To find the non-relativistic limit of Yukawa theory, we will use the scaling of the coordinates (\ref{sptmcontract}) and the fields as
\be{} \varphi\rightarrow \varphi,~ \phi \rightarrow \phi ,~ \chi \rightarrow \epsilon \chi. \ee
Using this limit on (\ref{eomyt1}), we get
\be{} \p^{i}\p_{i}\varphi=0 ,~ i\sigma^{i}\p_{i}\phi=0,~
i\p_{t}\phi + i\sigma^{i}\p_{i}\chi -g\phi\varphi =0. \ee
The scale-spin representation of GCA is determined by input from dynamics as was the case for the free theories. Here we are dealing with spinors $(\phi,\chi)$. For the Galilean Yukawa theory, the limit from the relativistic theory fixes the values of the constants to:
\be{} \lbrace (\D_1, \D_2), (a,b,\tb)\rbrace=\lbrace (1, \frac{3}{2}), (0,0,\frac{1}{2})\rbrace. \ee
We are now in a position to check for the invariance under GCA using (\ref{prosc}) and  (\ref{inftr}). The EOM are trivially invariant under $M_{l}^{(n)}$. Checking for the invariance under $L^{(n)}$, we have 
\be{} 
[L^{(n)},i\p_{t}\phi + i\sigma^{i}\p_{i}\chi -g\phi\varphi ]= in(n+1)t^{n-1}(\Delta_2 -\frac{3}{2})\phi+(1-\Delta_1)(n+1)t^{n}g\phi\varphi,\ee
\be{}
\p^{i}\p_{i}[L^{(n)},\varphi]=0,~ i\sigma^{i}\p_{i}[L^{(n)},\phi]=0.
 \ee
The EOM have Galilean conformal invariance in $D=4$.

\subsection{Electrodynamics with Scalar fields}
Now we will add matter to the free Maxwell theory and will try to understand the non-relativistic limit and its symmetries. The dynamics of scalar electrodynamics is given by the Lagrangian density: 
\bea{lsed}
\mathcal{L} = -\frac{1}{4} F_{\mu\nu}F^{\mu\nu}-(D_{\mu}\varphi)^{*}(D^{\mu}\varphi),
\eea 
where $F_{\mu\nu}= \p_{\mu}A_{\nu}-\p_{\nu}A_{\mu}$ is the electromagnetic field strength and $D_{\mu}=\p_{\mu}-ieA_{\mu}$ is the covariant derivative with $e$ being the electric charge.
The EOM are given as
\bea{sedeom} \p^{\mu}F_{\mu\nu}-ie\varphi^{*}(D_{\nu}\varphi) +ie(D_{\nu}\varphi)^{*}\varphi=0,~
\p^{\mu}(D_{\mu}\varphi)-ieA^{\mu}(D_{\mu}\varphi)=0.
\eea
It is obvious that the above equations are invariant under Poincare transformations (\ref{poin}). Checking the invariance of EOM under scale transformation (\ref{st}), we find:
\bes
\bea{}
\delta_{D}[ \p^{\mu}F_{\mu\nu}-ie\varphi^{*}(D^{\nu}\varphi) +ie (D_{\nu}\varphi)^{*}\varphi]= -2ie(\Delta_2 -1)[\varphi^{*}(D_{\nu}\varphi) - (D_{\nu}\varphi)^{*}\varphi]\non\\+(\Delta_1 -1)[\p^{\mu}F_{\mu\nu}-2e^2 A_{\nu}\varphi^{*}\varphi],\\
\delta_{D}[\p^{\mu}(D_{\mu}\varphi)-ieA^{\mu}(D_{\mu}\varphi)]= -ie(\Delta_1 -1)[\p^{\mu}
(A_{\mu}\varphi)+A^{\mu}(D_{\mu}\varphi)-ieA^{\mu}A_{\mu}\varphi].
\eea
\ees
By usual scaling arguments, the theory is seen to be invariant under dilatations in $D=4$.
Under special conformal transformations (\ref{sctt}):
\bea{} &&\delta^{\sigma}_{K}[\p^{\mu}F_{\mu\nu}-ie\varphi^{*}(D^{\nu}\varphi) +i(D_{\nu}\varphi)^{*}\varphi]= (2\Delta_1 +6-2D)F^{\sigma}_{~\nu}+(2\Delta_1 -2)[x^{\sigma}\p^{\mu}F_{\mu\nu}+\p^{\sigma}A_{\nu}\non\\&&~~~~~~~~~~~~~~~~~~~~~~~~~~~~~~~~~~~~~~~~~~~~~~~~~-\delta^{\sigma}_{\nu}\p^{\mu}A_{\mu}-2x^{\sigma}e^2 A_{\nu}\varphi^{*}\varphi]-ie(4\Delta_2 -4)x^{\sigma}[\varphi^{*}(D_{\nu}\varphi)\non\\&&~~~~~~~~~~~~~~~~~~~~~~~~~~~~~~~~~~~~~~~~~~~~~~~~~ - (D_{\nu}\varphi)^{*}\varphi],\non
\eea
\bea{}&&  \delta^{\sigma}_{K}[\p^{\mu}(D_{\mu}\varphi)-ieA^{\mu}(D_{\mu}\varphi)]= -2ie(\Delta_1 -1)[A^{\sigma}\varphi +x^{\sigma}\p^{\mu}(A_{\mu}\varphi)+x^{\sigma}A^{\mu}(D_{\mu}\varphi)-iex^{\sigma}A^{\mu}A_{\mu}\varphi]\non\\&&~~~~~~~~~~~~~~~~~~~~~~~~~~~~~~~~~~~~~~~+(4+4\Delta_2-2D)D^{\sigma}\varphi.\non\eea
The EOM are invariant under special conformal transformations in $D=4$, when we put in the values of $\D_1, \D_2$.

\subsection*{Galilean Scalar Electrodynamics}  
We will now move to the non-relativistic limit of Scalar Electrodynamics. To take the limit on the complex scalar field $\varphi$, we will decompose it into two real fields
\be{dcs} \varphi = \frac{1}{\sqrt{2}}(\psi_{1}+i\psi_{2}). \ee 
This decomposition and the subsequent scalings would lead to a non-trivial with non-zero interaction pieces. After decomposition, the relativistic equations (\ref{sedeom}) become
\bea{} \p^{\mu}F_{\mu\nu}+e[\psi_{1}\p_{\nu}\psi_{2}- \psi_{2}\p_{\nu}\psi_{1}-eA_{\nu}(\psi^{2}_{1}+\psi^{2}_{2})]=0,\label{sed1} \\
\p^{\mu}\p_{\mu}\psi_1 + e\p^{\mu}A_{\mu}\psi_2 +2eA_{\mu}\p^{\mu}\psi_2 -e^2 A^{\mu}A_{\mu}\psi_1=0, \label{sed2} \\
\p^{\mu}\p_{\mu}\psi_2 - e\p^{\mu}A_{\mu}\psi_1 -2eA_{\mu}\p^{\mu}\psi_1 -e^2 A^{\mu}A_{\mu}\psi_2=0. \eea  
To obtain the non-relativistic limits, we will scale $(\psi_1 ,\psi_2)$. In Sec[\ref{vfr}], we have seen that the gauge field can be scaled in two ways (\ref{sca}), even here we will have two limits of the non-relativistic limits for Galilean scalar electrodynamics.

\subsubsection*{Electric sector} 
In this limit, we will scale the scalar and vector potentials identically as done in the electric limit of Galiean Electrodynamics. The  scaling of the fields are given as
\bea{}\label{scesed1} A_{i} \rightarrow \epsilon A_{i}, \hspace{.2cm}
A_{t} \rightarrow A_{t}, \hspace{.2cm} \psi_{1}\rightarrow \epsilon \psi_{1}, \hspace{.2cm} \psi_{2}\rightarrow \frac{1}{\epsilon}\psi_{2}.\eea
The EOM are
\bes \label{esed1}
\bea{}
\p^{j}\p_{j} A_{i}-\p^{j}\p_{i} A_{j}+\p_{i}\p_{t} A_{t} +e[\psi_{1}\p_{i} \psi_{2}-\psi_{2}\p_{i} \psi_{1} -eA_{i}\psi^{2}_{2}]=0,\\
\p^{i}\p_{i}A_{t}-e^{2}A_{t}\psi^{2}_{2}=0, \hspace{.2cm} \p^{i}\p_{i}\psi_{2}=0,\\
\p^i\p_i \psi_1 -e\p_t A_t \psi_2 +e\p^i A_i \psi_2 -2eA_t \p_t \psi_2 +2eA_i \p^i \psi_2 =0.
\eea
\ees  
To make things more readable, we will denote the left hand side of the first EOM as $\mathcal{A}$, second as $\mathcal{B}$ and fourth as $\mathcal{C}$. Thus, the equations  in these notations become
\be{}
\mathcal{A}=0,~~ \p^{i}\p_{i}\psi_{2}=0, ~~\mathcal{B}=0,~~ \mathcal{C}=0.
\ee
Before going into the process of finding the invariance of the EOM, let us look into the remnants of gauge invariance in this limit. The gauge transformations for relativistic scalar electrodynamics are given by
\be{gsed}
\varphi(x)\rightarrow e^{ie\alpha(x)}\varphi(x) , ~~ A_{\mu}(x)\rightarrow A_{\mu}(x)+\p_{\mu}\alpha(x), \ee
where $\alpha(x)$ is an arbitrary function of spacetime. In terms of real scalar fields, the gauge transformation reads:
\bea{} \psi^{'}_{1} = \psi_{1}\text{cos}(e\alpha)- \psi_{2}\text{sin}(e\alpha),~
  \psi^{'}_{2} = \psi_{1}\text{sin}(e\alpha)+ \psi_{2}\text{cos}(e\alpha).\eea
The infinitesimal transformations are  
\bea{}\label{csed} \delta\psi_{1}(x)=-e\alpha\psi_{2}(x),~~
 \delta\psi_{2}(x)= e\alpha\psi_{1}(x).\eea  
Applying the scaling (\ref{scesed1}) to (\ref{gsed}) and checking the scaling for $\alpha$, we find that the scaling needs to be
\bea{} \alpha(x)\rightarrow \epsilon^{2}\alpha(x).\eea 
The non-relativistic version of gauge transformation in this limit reads:
\bea{} A_i (x) \rightarrow A_i(x)+ \partial_i \alpha(x), ~~~~
 A_t (x) \rightarrow A_t(x),\\
\psi_{1}(x) \rightarrow \psi_{1}(x)-e\alpha \psi_{2}(x), ~~~ \psi_{2}(x) \rightarrow \psi_{2}(x).\eea
The EOM (\ref{esed1}) are invariant under the new gauge transformations. 

Returning to the check of the invariance of EOM under scale and special conformal transformations, we find that the values of the constants which fix the details of the representation theory are required. These, as before, are taken to be the ones originating from the free parts. For Electric sector, the values of the constants are
\be{} \lbrace (\D_1, \D_2), (a,r,s) \rbrace=\lbrace (1, 1), (0,-1,0) \rbrace. \ee
The invariance under dilatation is given as:
\bes
\bea{}[D,\mathcal{B}] =-2e^{2}(\Delta_2 -1)A_{t}\psi^{2}_{2}, ~~~~~~ \p^{i}\p_{i}[D,\psi_{2}]=0,\\
~~[D,\mathcal{A}]= (\Delta_1 -1)(\p^{j}\p_{j} A_{i}-\p^{j}\p_{i} A_{j}+\p_{i}\p_{t} A_{t}-e^2A_i \psi^{2}_{2})\non\\+2e(\Delta_2 -1)( \psi_{1}\p_{i}\psi_{2}-\psi_{2}\p_{i}\psi_{1}
-eA_{i}\psi^{2}_{2}).   
\eea
\ees 
The invariance under special conformal transformations is
\bea{} [K_{l},\mathcal{A}] = 0,
~\p^{i}\p_{i}[K_{l},\psi_{2}]=0,
~[K_{l},\mathcal{B}] = 0,~ [K_l , \mathcal{C}]=0.\eea
Similary for $K$, we have
\bes
\bea{} && [K,\mathcal{B}] = -4te^{2}(\Delta_2 -1)A_{t}\psi^{2}_{2},~\p^{i}\p_{i}[K,\psi_{2}]=0,\\
&& [K,\mathcal{A}] = (3-D+\Delta_1)2t\p_{i}A_{t}
+2t[(\Delta_1 -1)(\p^{j}\p_{j} A_{i}-\p^{j}\p_{i} A_{j}+\p_{i}\p_{t} A_{t}-e^2A_i \psi^{2}_{2})\non\\ && \qquad \qquad +2e(\Delta_2 -1)( \psi_{1}\p_{i}\psi_{2}-\psi_{2}\p_{i}\psi_{1}
-eA_{i}\psi^{2}_{2})].\eea
\ees
Hence, we see that the EOM are invariant in $D=4$.
Let us now check the invariance under infinite dimesional GCA. The EOM are trivially invariant under $M_{l}^{(n)}$. Under $L^{(n)}$, we have
\bea{} [L^{(n)},\mathcal{B}] = -2(n+1)t^{n}e^{2}(\Delta_2 -1)A_{t}\psi^{2}_{2},
~\p^{i}\p_{i}[L^{(n)},\psi_{2}]=0,
\eea
\bea{}
 [L^{(n)},\mathcal{A}] = && (3-D+\Delta_1)n(n+1)t^{n-1}\p_{i}A_{t}
+(n+1)t^{n}[(\Delta_1 -1)\non\\&&~(\p^{j}\p_{j} A_{i}-\p^{j}\p_{i} A_{j}+\p_{i}\p_{t} A_{t}-e^2A_i \psi^{2}_{2})\non\\&&+2e(\Delta_2 -1)( \psi_{1}\p_{i}\psi_{2}-\psi_{2}\p_{i}\psi_{1}
-eA_{i}\psi^{2}_{2})],\eea
\bea{} [L^{(n)},\mathcal{C}]=&& (D-1-\Delta_1 -2\Delta_2)n(n+1)t^{n-1}eA_t \psi_2 \non\\ && -(\Delta_1 -1)et^n (n+1)[\p_t A_t \psi_2 -\p^{i}A_i \psi_2 +2A_t \p_t \psi_2 -2A_i \p^i \psi_2].\eea
The EOM are invariant under $M_{l}^{(n)}$ and $L^{(n)}$.
\subsubsection*{Magnetic sector}
In Magnetic sector, the scaling of the fields are given by
\bea{}\label{scmsed1} A_{i} \rightarrow  A_{i}, \hspace{.2cm}
A_{t} \rightarrow \epsilon A_{t}, \hspace{.2cm} \psi_{1}\rightarrow \psi_{1}, \hspace{.2cm} \psi_{2}\rightarrow \frac{1}{\epsilon}\psi_{2}.\eea
The EOM are
\bes \label{msed1}
\bea{}
\p^{i}\p_{i} A_{t}-\p^{i}\p_{t} A_{i}+e[\psi_{1}\p_{t} \psi_{2}-\psi_{2}\p_{t} \psi_{1} -eA_{t}\psi^{2}_{2}]=0,\\
\p^{j}\p_{j} A_{i}-\p^{j}\p_{i} A_{j}+e[\psi_{1}\p_{i} \psi_{2}-\psi_{2}\p_{i} \psi_{1} -eA_{i}\psi^{2}_{2}]=0,\\
\p^i \p_i \psi_1 +e\p^i A_i \psi_2 +2e A_i \p^i \psi_2 =0,\\
\p^{i}\p_{i}\psi_{2}=0.\eea
\ees
 We will denote the left hand side of the first EOM as $\mathcal{C}$, the second as $\mathcal{D}$ and the third as $\mathcal{E}$. Thus, the EOM in this notation become
\be{} \mathcal{C}=0, ~~\mathcal{D}=0,~~ \p^{i}\p_{i}\psi_{2}=0,~ \mathcal{E}=0.\ee
We will now look into the gauge transformations in this limit. The gauge transformation for magnetic sector can be found by taking
the scaling of the fields (\ref{scmsed1}) and $\alpha$ as
\be{} \alpha(x)\rightarrow \epsilon\alpha(x).\ee 
 Using the limit on (\ref{gsed}), the transformation becomes
\bea{} A_i (x) \rightarrow A_i(x)+ \partial_i \alpha(x), ~~~~
 A_t (x) \rightarrow A_t (x)+ \partial_t \alpha(x),\\
\psi_{1}(x) \rightarrow \psi_{1}(x)-e\alpha \psi_{2}(x), ~~~ \psi_{2}(x) \rightarrow \psi_{2}(x).\eea
The above transformation leave the EOM invariant.
We will now find the invariance of EOM under scale and special conformal transformations. The values of the constants are taken as
\be{} \lbrace (\D_1, \D_2), (a,r,s) \rbrace=\lbrace (1,1), (0,0,-1) \rbrace. \ee
The invariance under scale transformation is given by
\bes
\bea{} &&[D,\mathcal{C}] = (\Delta_1 -1)[ (\p^{i}\p_{i} A_{t}-\p^{i}\p_{t} A_{i}-e^2 A_t \psi^{2}_2)]\non\\&&\qquad \qquad+(2\Delta_2 -2)e [\psi_{1}\p_{t}\psi_{2}-\psi_{2}\p_{t}\psi_{1}
-eA_{t}\psi^{2}_{2}],\\
&& [D,\mathcal{D}]=  (\Delta_1 -1)(\p^{j}\p_{j} A_{i}-\p^{j}\p_{i} A_{j}-e^2 A_i \psi^{2}_{2})\non\\&&\qquad \qquad +e(2\Delta_2 -2)[ \psi_{1}\p_{i}\psi_{2}-\psi_{2}\p_{i}\psi_{1}-eA_{i}\psi^{2}_{2}],\\
&& \p^{i}\p_{i}[D,\psi_{2}]=0.    
\eea
\ees 
Checking the invariance under $K_l$:
\bea{} [K_{l},\mathcal{C}] = 0,
~\p^{i}\p_{i}[K_{l},\psi_{2}]=0,
~[K_{l},\mathcal{D}] = 0,~[K_l,\mathcal{E}]=0.\eea
In a similar fashion, under $K$:
\bes
\bea{} [K,\mathcal{C}]= 2(\Delta_1 -1)[t (\p^{i}\p_{i} A_{t}-\p^{i}\p_{t} A_{i}-e^2 A_t \psi^{2}_2)-\p^{i}A_{i}]\non\\+(4\Delta_2 -4)et [\psi_{1}\p_{t}\psi_{2}-\psi_{2}\p_{t}\psi_{1}
-eA_{t}\psi^{2}_{2}],\\
~[K,\mathcal{D}]= 2t(\Delta_1 -1)(\p^{j}\p_{j} A_{i}-\p^{j}\p_{i} A_{j}-e^2 A_i \psi^{2}_{2})\non\\+2et(2\Delta_2 -2)[ \psi_{1}\p_{i}\psi_{2}-\psi_{2}\p_{i}\psi_{1}-eA_{i}\psi^{2}_{2}],\\
\p^{i}\p_{i}[K,\psi_{2}]=0.
\eea
\ees 
The EOM are invariant under finite GCA in $D=4$. We will now check the invariance under infinite extention of GCA. It can be seen that the EOM are trivially invariant under $M_{l}^{(n)}$. Under $L^{(n)}$, we have
\bea{} [L^{(n)},\mathcal{C}] = (n+1)(\Delta_1 -1)[t^n (\p^{i}\p_{i} A_{t}-\p^{i}\p_{t} A_{i}-e^2 A_t \psi^{2}_2)-nt^{n-1}\p^{i}A_{i}]\non\\+(2\Delta_2 -2)(n+1)et^n [\psi_{1}\p_{t}\psi_{2}-\psi_{2}\p_{t}\psi_{1}
-eA_{t}\psi^{2}_{2}],\\
~[L^{(n)},\mathcal{D}] = (n+1)t^n(\Delta_1 -1)(\p^{j}\p_{j} A_{i}-\p^{j}\p_{i} A_{j}-e^2 A_i \psi^{2}_{2})\non\\+e(n+1)t^{n}(2\Delta_2 -2)[ \psi_{1}\p_{i}\psi_{2}-\psi_{2}\p_{i}\psi_{1}-eA_{i}\psi^{2}_{2}],\eea
\bea{} [L^{(n)},\mathcal{E}]=(\Delta_1-1) e(n+1)t^n [\p^i A_i \psi_2 +2A_i \p^i \psi_2],\eea
along with $\p^{i}\p_{i}[L^{(n)},\psi_{2}]=0$.
The EOM are invariant in $D=4$ dimension.

\subsection{Electrodynamics with Fermions}
We now consider the non relativistic limit of $U(1)$ gauge fields coupled to fermions. The Lagrangian density for the relativistic case is
\bea{} \mathcal{L}=-\frac{1}{4}F_{\mu\nu}F^{\mu\nu} + i\bar{\Psi}\gamma^{\mu}D_{\mu}\Psi, \eea
where $F_{\mu\nu}= \p_{\mu}A_{\nu}-\p_{\nu}A_{\mu}$ is the electromagnetic field strength and $D_{\mu}=\p_{\mu}-ieA_{\mu}$ is the covariant derivative with $e$ being the electric charge.  The corresponding EOM are
\be{spedeom} \p_{\mu}F^{\mu\nu}+e\bar{\Psi}\gamma^{\nu}\Psi=0,~
i\gamma^{\nu}D_{\nu}\Psi=0.\ee
We can check that the Lagrangian density and EOM  are invariant under the finite local gauge transformations 
\bea{} 
&& \Psi(x)\rightarrow \Psi'(x)=e^{ie\theta(x)}\Psi,~\bar{\Psi}(x)\rightarrow \bar{\Psi}'(x)=\bar{\Psi}(x)e^{-ie\theta(x)},\\\non
&& A_{\mu}(x)\rightarrow A_{\mu}'(x)=A_{\mu}(x)+\p_{\mu}\theta(x)\eea
where $\theta(x)$ is the real and finite local parameter of transformation.\\ 
Next, we will check for invariance of (\ref{spedeom}) under scale (\ref{st}) and special conformal transformations (\ref{sctt}). Under scale transformation,
\bea{} \delta_{D}[\p_{\mu}F^{\mu\nu}+e\bar{\Psi}
\gamma^{\nu}\Psi] =(\Delta_{1} -1)\p_{\mu}F^{\mu\nu}+(2\Delta_{2}-3)e\bar{\Psi}
\gamma^{\nu}\Psi, \\
\delta_{D}[ i\gamma^{\mu}\p_{\mu}\Psi + e\gamma^{\mu}A_{\mu}\Psi] =(\Delta_{1}-1)e\gamma^{\tau}A_{\tau}\Psi.\eea
The equation is invariant for $\Delta_{1}=1$ and $\Delta_{2}=\frac{3}{2}$.
Under special conformal transformations, we have
\bea{} \delta_{K}^{\sigma}[\p_{\mu}F^{\mu\nu}+e\bar{\Psi}
\gamma^{\nu}\Psi]=
 2(\Delta_{1}-1)(x^{\sigma}\p_\mu F^{\mu\nu}+F^{\sigma\nu}+\p^{\sigma}A^{\nu}
 -\eta^{\nu\sigma}\p_{\mu}A^{\mu})\non\\+(4\Delta_{2}-6)x^{\sigma}e\bar{\Psi}\gamma^{\nu}\Psi,
\\
 \delta_{K}^{\sigma}[i\gamma^{\mu}\p_{\mu}\Psi + e\gamma^{\mu} A_{\mu}\Psi] =2(\Delta_{1}-1)x^{\sigma}e\gamma^{\tau}A_{\tau}\Psi +i(2\Delta_{2}
-3)\gamma^{\sigma}\Psi.\eea    
The EOM are invariant, since $\Delta_{1}=1$ and $\Delta_{2}=\frac{3}{2}$.
\subsection*{Galilean Electrodynamics with Fermions}\label{gseds}
To take the non-relativistic limit, we first decompose the Dirac spinor into two components spinors as in Sec[\ref{gdft}]. By doing so, the relativistic equations (\ref{spedeom}) become
\bea{} \p^{i}F_{it}-e(\phi^{\dagger}\phi+\chi^{\dagger}\chi)=0,~
-\p_{t}F_{t}^{~j}+\p_{i}F^{ij}+e(\phi^{\dagger}\sigma^{j}\chi
+\chi^{\dagger}\sigma^{j}\phi)=0,\label{gedf}\\
 (i\p_{t}+eA_{t})\phi +(i\sigma^{i}\p_{i}+e\sigma^{i}A_{i})\chi=0 ,~
(i\sigma^{i}\p_{i}+e\sigma^{i}A_{i})\phi+ (i\p_{t}+eA_{t})\chi=0. \label{gedf1}
\eea
We know that the gauge field can be scaled in two ways (\ref{sca}). Similarly, we can also scale the spinors $(\phi,\chi)$ in two ways in accordance to the limit of gauge field under consideration. 
\subsubsection*{Electric Sector}
In this limit, we take the scaling of the fields as
\bea{esl} A_{i} \rightarrow \epsilon A_{i},~
A_{t} \rightarrow A_{t},~\phi\rightarrow \frac{1}{\epsilon} \phi,~\chi\rightarrow \chi.\eea 
The EOM can be found by using (\ref{sptmcontract}) and (\ref{esl}) on (\ref{gedf}) and \refb{gedf1}:
\bes \label{esse}
\bea{} \p_{i}\p^{i}A^{j}-\p^{i}\p^{j}A_{i}+\p_{t}\p^{j}A_{t}
+e(\phi^{\dagger}\sigma^{j}\chi+\chi^{\dagger}
\sigma^{j}\phi)=0,\\
\p^{i}\p_{i}A_{t}-e\phi^{\dagger}\phi=0,~
(i\p_{t}+eA_{t})\phi +i\sigma^{i}\p_{i}\chi=0,~ i\sigma^{i}\p_{i}\phi=0.\label{eomesf}
\eea
\ees
 We denote the first EOM as $\mathcal{E}$, second as $\mathcal{F}$ and third as $\mathcal{G}$.
 From now on, we will only show the invariance of the EOM under infinite extension of GCA. The invariance under scale and special conformal transformation can be read directly from it. \\ \\To check the symmetries of the EOM, we need the set of
constants $(r, s,b,\tb)$. They are given by
\be{}\lbrace r, s,b,\tb\rbrace=\lbrace -1,0,0,\frac{1}{2}\rbrace. \ee
Along with this, we also need the dilation eigenvalues which are 
\be{}
\D_1= 1, \quad \D_2= \frac{3}{2}.
\ee
These remain fixed irrespective of the particular sector of the theory we consider. The invariance can be trivially checked under $M_{l}^{(n)}$.
Checking for the invariance under $L
^{(n)}$, we have
\bea{} [L^{(n)},\mathcal{E}]= (n+1)(\Delta_{1}-1)[t^n (\p_{i}\p^{i}A_{j}-\p^{i}\p_{j}A_{i}+\p_{t}\p_{j}A_{t})
+nt^{n-1}\p_{j}A_{t}]\non\\+(2
\Delta_{2}-3)t^n(n+1) e(\phi^{\dagger}\sigma^{j}\chi+\chi^{\dagger}\sigma^{j}
\phi).\eea 
Similarly, we have
\bea{} [L^{(n)},\mathcal{F}]= (n+1)t^{n}[(\Delta_{1}-1)\p^{i}\p_{i}A_{t}-(2\Delta_{2}-3)e\phi^{\dagger}\phi]
,~i\sigma^{i}\p_{i}[L_{n},\phi]=0, \\
~[L^{(n)},\mathcal{G}]=
(n+1)[(\Delta_{1}-1)t^n eA_{t}\phi +int^{n-1}(\Delta_{2}-\frac{3}{2})\phi]. \eea
The EOM are invariant under the infinite modes of GCA in $D=4$. 
\subsubsection*{Magnetic Sector}
The fields in this limit are scaled as 
\be{} A_{i} \rightarrow  A_{i},~
A_{t} \rightarrow \epsilon A_{t},~\phi\rightarrow \epsilon \phi,~\chi\rightarrow \chi\ee
with EOM given as
\be{msse} \p^{i}\p_{i}A_{t} - \p^{i}\p_{t} A_{i} =0,~~ \p^{i}\p_{i}A_{j} - \p^{i}\p_{j} A_{i} =0,~~
i\sigma^{i}\p_{i}\chi =0 ,~~ i\sigma^{i}\p_{i}\phi +i\p_{t}\chi=0.\ee
We see that this is an uninteresting sector, which does not exhibit any interactions in the EOM. It just correspond to $U (1)$ magnetic sector and a free Galilean Dirac theory. The invariance is quite trivial and follows from our previous analysis.

\section{Interacting Galilean field theories II}

\subsection{Pure Yang-Mills Theory} \label{rgym}
Till now, we have applied our analysis to free theories and Electrodynamics coupled to matter fields. In each case, we found that the non-relativistic version exhibits infinite enhanced Galilean conformal symmetries in appropriate dimension. Next obvious step will be to look into Yang-Mills theories. This was considered in \cite{Bagchi:2015qcw}. Here we briefly review our construction. When one considers the simplest of the Yang-Mills theories, the $SU(2)$ theory, there are already interesting features that emerge in the non-relativistic case. In going from $U(1)$ to $SU(2)$, the first non-trivial aspect of the generalisation is the existence of skewed limits. The presence of three different gauge fields leads to four distinct limits instead of two in the $U(1)$ theory. 

In \cite{Bagchi:2015qcw}, we performed a complete generalisation of this analysis to Yang-Mills theories with arbitrary gauge groups. The original Lorentzian gauge field 1-forms $A= A_a T^a$ took values in a semi-simple Lie-algebra $ \mathfrak{g}$ spanned by $T^a$, with $f^{abc}$ as structure constants. A total of $\mathfrak{D}+1$ distinct Galilean limits of the gauge theory emerged, if the dimension of the vector space $ \mathfrak{g}$ was taken to be $\mathfrak{D}$. To each of these limit sectors, we assigned one of the following $\mathfrak{D}$ dimensional vectors:
\bea{}
\Xi_{(p)} = (\underbrace{0,0,\dots , 0}_{\mathfrak{D}-p}, \underbrace{1,1,\dots , 1}_{p}) \qquad ~~ p=0, \dots, \mathfrak{D} 
\eea
where we denoted the $a^{\mathrm{th}}$ component of $\Xi_{(p)}$ as $\Xi_{(p)}^a$, which can take values $0$ or $1$. Our next step was to concentrate on a given sector, let's say the Galilean $p_0^{\mathrm{th}}$ sector, $p_0$ however is arbitrary. In this sector, gauge fields was found from the relativistic gauge field by the following contraction:
\bea{sgf} 
A^a_t \longrightarrow \frac{\epsilon}{1+\epsilon-\Xi^a_{(p_0)}}A^a_t,		~~~~ 	A^a_i \longrightarrow \frac{\epsilon}{\epsilon+\Xi^a_{(p_0)}}A^a_i.
\eea
If $\Xi_{(p_0)}^a=1$, then the scaling for the (Galilean) scalar and vector parts of the corresponding gauge field component were given as
\be{ymcontra1}
A^a_t \rightarrow  A^a_t, \,  \, A^a_i \rightarrow  \e A^a_i, 
\ee
and if $\Xi_{(p_0)}^a=0$, the scaling on the gauge field became:
\be{}
A^a_t \rightarrow \e A^a_t, \,  \, A^a_i \rightarrow  A^a_i.
\ee
A more convenient notation for the gauge fields was introduced. If the index $a$ was in the range $1 \leq a \leq \mathfrak{D}-p_0$, it was denoted by capital Romans letters $I,J \dots$ and if it was in the range $ \mathfrak{D}-p_0+1 \leq a \leq \mathfrak{D}$ then by Greeks letters $ \alpha, \beta \dots$ . In these notations, \eqref{ymcontra1} become respectively:
\bea{ymcontra}
  A^{\alpha}_t \rightarrow  A^{\alpha}_t,~   A^{\alpha}_i \rightarrow  \e A^{\alpha}_i ~~
\mbox{and }~~ A^I_t \rightarrow \e A^I_t,~  A^I_i \rightarrow  A^I_i.
\eea
The two extreme sectors, $p_0=\mathfrak{D}$ and $p_0= 0$ were the `vanilla' Electric and the Magnetic limits. In between, there were a plethora of skewed limits. The EOM in the $\mathfrak{D}+1$ distinct Galilean sectors were written by looking at the contraction rules prescribed by the $ \Xi_{(p_0)}$ vector (\ref{ymcontra}). These are described below.
\paragraph{Case 1: $\mathfrak{D}-p_0+1 \leq a \leq \mathfrak{D}$}
The scalar equation is:
\bea{gymeom3}
\p^i \p_i A_t^{ \alpha}=0.
\eea
The spacetime vector one is:
\bea{gymeom4}
&& \p_t \p_j A_t^{\alpha} + \p^i \left( \p_i A_j^{\alpha} - \p_j A_i^{\alpha} \right) \non\\ &&+g \bigg [f^{ \alpha}{}_{\beta \gamma} A_t^{\beta} \p_j A_t^{ \gamma} +  f ^{\alpha}{}_{ J K} \p^i \left(   A^J_i A^K_j \right) + f^{ \alpha}{}_{ J K } A^{iJ} \left( \p_i A^K_j - \p_j A^K_i\right) \bigg]=0.
\eea

\subsection*{Case 2: $1 \leq a \leq \mathfrak{D}-p_0$}
Scalar equation:
\bea{gymeom1}
\p^i \p_i A_t^I -\p^i \p_t A_i^I + g f^{I}{}_{J\alpha} \left[\p^i \left(  A_i^J A_t^{ \alpha}\right) + A_i^J \p^i A_t^{\alpha}  \right]=0.
\eea
Vector equation:
\bea{gymeom2}
\p ^i (\p_i A_j^I - \p_j A_i^I)=0.
\eea
We found that the invariance of the EOM under the infinite dimensional Galilean conformal algebra holds for $D=4$ (and not other dimensions). For more details, we recommend the reader look at \citep{Bagchi:2015qcw}.

\subsection{Yang-Mills with Fermions}
We will now extend our analysis  to Yang-Mills coupled with fermions. Let us start with Lagrangian density of the theory
\bea{}\label{spnyml} \mathcal{L}=-\frac{1}{4}F_{\mu\nu}^{a}F^{\mu\nu a} + i\bar{\psi}_m \gamma^{\mu}(D_{\mu}\psi)_m ,\eea
where $F_{\mu\nu}^{a}= \partial_{\mu} A_{\nu}^{a} - \partial_\nu A_{\mu}^{a}+gf^{abc}A^{b}_{\mu}A^{c}_{\nu}$ is the non abelian field tensor and $ D_\mu \equiv \partial_\mu-igT^{a}A_{\mu}^{a}$ is the non abelian gauge covariant derivative with $g$ being the coupling constant. The label $a$ is the colour index, $m$ is an internal symmetry index and $f^{abc}$ are the structure constants of the underlying gauge group with generators following the algebra $[T^{a},T^{b}]=if^{abc}T^{c}$.
The EOM for this theory are 
\be{}\label{eomsymm}  \p_{\mu}F^{\mu \nu a}+gf^{abc}A^{b}_{\mu}F^{\mu\nu c}+g\bar{\psi}_{m} \gamma^{\nu}T^{a}_{mn}\psi_{n}=0,~~
i\gamma^{\nu}(D_{\nu}\psi)_n =0.  \ee
We know that the Lagrangian density and EOM are invariant under the following finite gauge transformations 
\bea{} 
\Psi_m(x)\rightarrow \Psi'_m(x)=(e^{i\theta^{a}T^a}\Psi)_m, \quad A^{a}_{\mu}(x)\rightarrow A^{a'}_{\mu}(x)=A^{a}_{\mu}+\frac{1}{g}\p_{\mu}\theta^{a}+f^{abc}A^{b}_{\mu}\theta^{c},\eea
where $\theta(x)$ is the real and finite local parameter of transformation. \\
To examine the symmetry of the EOM, we require the transformation of field strength under conformal transformations. They are given as
\bea{} \delta_{D}F^{\mu\nu a}(x)=(x^{\tau}\partial_{\tau}+\Delta_{1}+1)F^{\mu\nu a}+(\Delta_1 -1)gf^{abc} A^{\mu b}A^{\nu c},\non\eea
\bea{}
\delta^{\sigma}_{K}F^{\mu\nu a} =
(2x^{\sigma}x^{\tau}\partial_{\tau}- x^{2}\partial^{\sigma})F^{\mu\nu a}+2(\Delta_1 +1)x^{\sigma}F^{\mu\nu a}+2\eta^{\sigma\mu}
x_{\tau}F^{\tau\nu a}
+2\eta^{\sigma\nu}x_{\tau}F^{\mu\tau a}\non\\
-2x^{\mu}F^{\sigma\nu a}
-2x^{\nu}F^{\mu\sigma a}+2(\Delta_1 -1)[\eta^{\sigma\mu}A^{\nu a}-\eta^{\sigma\nu}A^{\mu a}
+gf^{abc}x^{\sigma}A^{\mu b}A^{\nu c}].\non
\eea
We will now check the invariance of (\ref{eomsymm}) under scale (\ref{st}) and special conformal transformations (\ref{sctt}). Under scale transformation,
\bea{}
\delta_{D}[ \p_{\mu}F^{\mu \nu a}+gf^{abc}A^{b}_{\mu}F^{\mu\nu c}+g\bar{\psi}_m \gamma^{\nu}T^{a}_{mn}\psi_n]= (\Delta_1 -1)[\p_{\mu}F^{\mu\nu a}+gf^{abc}\lbrace\p_{\mu}(A^{\mu b}A^{\nu c})\non\\+2A_{\mu}^b F^{\mu\nu c}+gf^{cde}A^{\mu b}A^{d}_{\mu}A^{\nu e}\rbrace]+(2\Delta_2 -3)g\bar{\psi}_m \gamma^{\nu}T^{a}_{mn}\psi_n ,  \non\eea
\bea{}
\delta_{D}[i\gamma^{\nu}(D_{\nu}\psi)_m ]= g (\Delta_1 -1)\gamma^{\mu}T^{a}_{mn}
A^{a}_{\mu}\psi_{n}.
\eea
Under special conformal transformations, we have
\bea{} \delta^{\sigma}_{K}[\p_{\mu}F^{\mu \nu a}+gf^{abc}A^{b}_{\mu}F^{\mu\nu c}+g\bar{\psi}_m \gamma^{\nu}T^{a}_{mn}\psi_n]= (2\Delta_1 -2)[(\p^{\sigma}A^{\nu a}-\eta^{\sigma\nu}\p_{\mu}A^{\mu a})+F^{\sigma\nu a}\non\\+gf^{abc}\lbrace \p_{\mu}(x^{\sigma}A^{\mu b}A^{\nu c})+(A^{\sigma b}A^{\nu c}-\eta^{\nu\sigma}A^{b}_{\mu}A^{\mu c})\non\\+ gf^{cde}x^{\sigma}A^{b}_{\mu}A^{\mu d}A^{\nu e}+2x^{\sigma}A^{b}_{\mu}F^{\mu\nu c}\rbrace +x^{\sigma}\p_\mu F^{\mu\nu a}]\non\\+
(4\Delta_2 -6)gx^{\sigma}\bar{\psi}_{m}\gamma^{\nu}T^{a}_{mn}\psi_{n},\non\eea
\bea{}  \delta^{\sigma}_{K}[i\gamma^{\nu}(D_{\nu}\psi)_m]= 2g(\Delta_1 -1)x^{\sigma}\gamma^{\mu}T^{a}_{mn}A_{\mu}^{a}\psi_{n}
+2i(\Delta_2 -\frac{3}{2})\gamma^{\sigma}\psi_{m}.\eea
Hence, the EOM are conformally invariant.
\bigskip

\subsection{Galilean $SU(2)$ Yang-Mills with Fermions}
In this section, we consider the non-relativistic limit of $SU(2)$ case of Yang-Mills coupled with fermions. The details of a more general analysis for any gauge group is left for future work, as the present case itself is rather involved and there are aspects of it that we don't fully appreciate yet. 

Under the decomposition of Dirac fermion (\ref{compbr}), the relativistic EOM are  
\bea{gsymeomc} 
&& {\hspace{-0.5cm}} \p_{t}F_{t}^{~ja}-\p_{i}F^{ija} + gf^{abc}(A^{b}_{t}F_{t}^{~jc} - A^{b}_{i}F^{ijc}) - g(\phi^{\dagger}_{m} \sigma^{j}T^{a}_{mn}\chi_{n} +\chi^{\dagger}_{m} \sigma^{j}T^{a}_{mn}\phi_{n})=0,\\
&& \p_{i}F^{ia}_{~t}+gf^{abc}A^{b}_{i}F^{ic}_{~t} -g(\phi^{\dagger}_m T^{a}_{mn}\phi_{n} +\chi^{\dagger}_{m} T^{a}_{mn}\chi_{n})=0,\\
&& i\p_{t}\phi_m +gT^{a}_{mn}A_{t}^{a}\phi_{n} +i\sigma^{i}\p_{i}\chi_{m} +g\sigma^{i}T^{a}_{mn}A_{i}^a \chi_{n} =0,\\
 && i\p_{t}\chi_{m}+gT^{a}_{mn}A_{t}^{a}\chi_{n} +i\sigma^{i}\p_{i}\phi_{m} +g\sigma^{i}T^{a}_{mn}A_{i}^a \phi_{n} =0.
\eea 
We are in the $SU(2)$ theory and hence $f^{abc} = \e^{abc}$ and $T^a = \s^a/2$. 

Next, as we have done throughout the course of this work, we will scale our fields along with the spacetime contraction to get the non-relativistic limits of this theory. In Sec[\ref{rgym}], we mentioned that the scaling of the gauge fields give rise to four distinct set of limits. We need to specify the scaling of the spinors. The Dirac spinors $(\Psi_1 , \Psi_2)$ belongs to $SU(2)$ group and can be decomposed into two copies of two component spinors ($\phi,\chi$). Each spinor can be assigned an general scaling, given as
\be{}\phi_1 \rightarrow \epsilon^{\a_1}\phi_1, \, \phi_2 \rightarrow \epsilon^{\a_2}\phi_2,~\chi_1 \rightarrow \epsilon^{\b_1}\chi_1,~\chi_2 \rightarrow \epsilon^{\b_2}\chi_2 ,\ee where $(\a_1,\a_2,\b_1,\b_2)$ are constant parameters.

We then take the relativistic EOM and plug these scalings together with the scaling of gauge fields. There are some constraints on these parameters that arise naturally. A class of such constraints arise as we demand that the non-relativistic EOM should reproduce the EOM of the pure SU(2) GYM when the matter fields are turned off and likewise, should reduce to the Galilean version of the Dirac equation when the gauge fields are switched off. We also wish to look at those sets of equations which lead to interactions between the matter fields and the gauge fields. The detailed calculations for the various sectors are outlined in Appendix \ref{ap3}. 

Here we are interested in studying the symmetries of the EOM and for this we shall pick a representative element of the set considered in Appendix \ref{ap3}, where $\a_1=0,\a_2=0,\b_1=1,\b_2=1$. For all the other sets, the analysis is very similar and the same symmetry structures emerge.  

 \subsubsection*{EEE case:}
The scaling of the gauge fields and spinors are given by
 \be{} A_{i}^{a} \rightarrow \epsilon A_{i}^{a},~
A_{t}^{a} \rightarrow A_{t}^{a},~\phi_{1,2} \rightarrow  \phi_{1,2} ,~\chi_{1,2} \rightarrow \epsilon\chi_{1,2}. \ee
We apply the above scalings on the EOM (\ref{gsymeomc}) to obtain the EOM for this limit:
\bes
\bea{}
\p_t\p^j A_{t}^{a}+\p^{i}(\p_i A^{ja}-\p^j A_{i}^{a})+gf^{abc}A^{b}_{t}\p^{j}A_{t}^{c}=0,\label{118eq1}
\\ i\p_t\phi_{1} +gT^{a}_{11}A^{a}_{t}\phi_1 +gT^{a}_{12}A^{a}_{t}\phi_2 +i\sigma^{i}\p_{i}\chi_{1}=0,\label{118eq2}\\ i\p_t\phi_{2} +gT^{a}_{21}A^{a}_{t}\phi_1 +gT^{a}_{22}A^{a}_{t}\phi_2 +i\sigma^{i}\p_{i}\chi_{2}=0, \label{118eq3}\\
\p_i\p^i A_{t}^{a}=0,~ i\sigma^{i}\p_{i}\phi_{1,2}=0.
\eea
\ees
We now look at the invariance of the equations under $L^{(n)}$'s. The required input from dynamics is 
\be{} \lbrace r^a,s^{a},b_{1},b_{2},\tb_{1},\tb_{2}\rbrace =\lbrace -1,0,0,0,\frac{1}{2},\frac{1}{2}\rbrace. \non\ee
All the EOM are trivially invariant under $M^{(n)}_{l}$. The invariance under $L^{(n)}$ is given by 
\bea{} [L_n, (\ref{118eq1})] = (n+1)(\Delta_1 -1)[nt^{n-1}\p_j A^{a}_{t} +t^{n}gf^{abc}A^{b}_{t}\p^{j}A_{t}^{c}],~~~~~~~~~~~~~~~~~\\
~[L_n,(\ref{118eq2})] =in(n+1)t^{n-1}(\Delta_2 -\frac{3}{2})\phi_{1} ~~~~~~~~~~~~~~~~~~~~~~~~~~~~~~~~~~~~~~~~~~~\cr +(\Delta_1 -1)t^n (n+1)(gT^{a}_{11}A^{a}_{t}\phi_1 +gT^{a}_{12}A^{a}_{t}\phi_2),\\
~[L_n,(\ref{118eq3})] =in(n+1)t^{n-1}(\Delta_2 -\frac{3}{2})\phi_{2} ~~~~~~~~~~~~~~~~~~~~~~~~~~~~~~~~~~~~~~~~~~~\cr +(\Delta_1 -1)t^n (n+1)(gT^{a}_{21}A^{a}_{t}\phi_1 +gT^{a}_{22}A^{a}_{t}\phi_2),\\
~[L_n,\p_i\p^i A_{t}^{a}] =0,~[L_n,i\sigma^{i}\p_{i}\phi_{1,2}] =0.~~~~~~~~~~~~~~~~~~~~~~~~~~~~~~~~~~~~~~~~~~~~~~~~~
\eea 
The equation numbers in the commutators just mean that we are looking at the action of $L_n$ on the left hand side of the respective equations. The EOM come out to be invariant under $L_{n}$.
\subsubsection*{EEM case:}
The fields in EEM limit scales as
\be{} A_{i}^{1,2} \rightarrow \epsilon A_{i}^{1,2},~
A_{t}^{1,2} \rightarrow A_{t}^{1,2},~A_{i}^{3} \rightarrow  A_{i}^{3},~
A_{t}^{3} \rightarrow \epsilon A_{t}^{3},~\phi_{1,2} \rightarrow \phi_{1,2} ,~\chi_{1,2} \rightarrow \epsilon \chi_{1,2}. \ee
The EOM in this limit are given as
\bes
\bea{} \p_t\p^j A_{t}^{(1,2)}+\p_{i}(\p^i A^{j(1,2)}-\p^j A^{i(1,2)})=0,\\
\p_i (\p^i A^{j3}-\p^j A^{i3})=0,~ \p_i (\p^i A_{t}^{3}-\p_t A^{i3})=0,\\ \label{119eq1} i\p_t\phi_{1} +gT^{1}_{12}A^{1}_{t}\phi_2 +gT^{2}_{12}A^{2}_{t}\phi_2 +i\sigma^{i}\p_{i}\chi_{1}=0,\\\label{119eq2} i\p_t\phi_{2} +gT^{1}_{21}A^{1}_{t}\phi_1 +gT^{2}_{21}A^{2}_{t}\phi_1 +i\sigma^{i}\p_{i}\chi_{2}=0,\\
\p_i\p^i A_{t}^{(1,2)}=0,~ i\sigma^{i}\p_{i}\phi_{1,2}=0.
\eea
\ees 
Again input from dynamics fixes the values of the coefficients as  
\be{} \lbrace (r^1,s^{1}),(r^2,s^{2}),(r^3,s^{3}),b_{1},b_2,\tb_{1},\tb_2\rbrace =\lbrace (-1,0),(-1,0),(0,-1),0,0,\frac{1}{2},\frac{1}{2}\rbrace. \non\ee
All the EOM are trivially invariant under $M^{(n)}_{l}$. The invariance under $L^{(n)}$ is given by 
\bea{}~[L_n,(\ref{119eq1})] =in(n+1)t^{n-1}(\Delta_2 -\frac{3}{2})\phi_{1} ~~~~~~~~~~~~~~~~~~~~~~~~~~~~~~~~~~~~~~~~~~~\cr +(\Delta_1 -1)t^n (n+1)(gT^{1}_{12}A^{1}_{t}\phi_2 +gT^{2}_{12}A^{2}_{t}\phi_2),\\
~[L_n,(\ref{119eq2})] =in(n+1)t^{n-1}(\Delta_2 -\frac{3}{2})\phi_{2} ~~~~~~~~~~~~~~~~~~~~~~~~~~~~~~~~~~~~~~~~~~~\cr +(\Delta_1 -1)t^n (n+1)(gT^{1}_{21}A^{1}_{t}\phi_1 +gT^{2}_{21}A^{2}_{t}\phi_1),\\
~[L_n,\p_t\p^j A_{t}^{(1,2)}+\p_{i}(\p^i A^{j(1,2)}-\p^j A^{i(1,2)})] =n(n+1)t^{n-1}(\Delta_1 -1)\p_j A^{1,2}_{t},\\
~[L_n, \p_i (\p^i A_{t}^{3}-\p_t A^{i3})]=-n(n+1)t^{n-1}(\Delta_1 -1)\p^{i}A^{3}_{i},~[L_n,\p_i\p^i A_{t}^{(1,2)}]=0,\\
~[L_n,\p_i (\p^i A^{j3}-\p^j A^{i3})]=0,~[L_n,i\sigma^{i}\p_{i}\phi_{1,2}]=0.
\eea
The EOMs are invariant under $L_{n}$.
\subsubsection*{EMM case:}
The fields in EMM limit scales as
\be{} A_{i}^{1} \rightarrow \epsilon A_{i}^{1},~
A_{t}^{1} \rightarrow A_{t}^{1},~A_{i}^{2,3} \rightarrow  A_{i}^{2,3},~
A_{t}^{2,3} \rightarrow \epsilon A_{t}^{2,3},~\phi_{1,2} \rightarrow \phi_{1,2} ,~\chi_{1,2} \rightarrow \epsilon \chi_{1,2}. \ee
The EOM for this limit are given as 
\bes
\bea{}\label{113eq1}\non \p_t\p^j A_{t}^{1}+\p_{i}(\p^i A^{j1}-\p^j A^{i1}+gA^{i2}A^{j3}-gA^{i3}A^{j2})~~~~~~~~~~~~~~~~~~~~~~~~~~~\\+gA^{2}_{i}(\p^i A^{j3}-\p^j A^{i3})-gA^{3}_{i}(\p^i A^{j2}-\p^j A^{i2})=0,\\
\p_i (\p^i A^{j(2,3)}-\p^j A^{i(2,3)})=0,~\p_i\p^i A_{t}^{1}=0,~ i\sigma^{i}\p_{i}\phi_{1,2}=0,\\\label{113eq2} \p_i (\p^i A_{t}^{2}-\p_t A^{i2}+gA^{i3}A_{t}^{1})+gA^{3}_{i}\p^{i}A_{t}^{1}=0,\\\label{113eq3} \p_i (\p^i A_{t}^{3}-\p_t A^{i3}-gA^{i2}A_{t}^{1})-gA^{2}_{i}\p^{i}A_{t}^{1}=0,\\\label{113eq4} i\p_t\phi_{1} +gT^{1}_{12}A^{1}_{t}\phi_2 +i\sigma^{i}\p_{i}\chi_{1}=0,\\\label{113eq5} i\p_t\phi_{2} +gT^{1}_{21}A^{1}_{t}\phi_1 +i\sigma^{i}\p_{i}\chi_{2}=0.
\eea
\ees
The values of the coefficients fixed by dynamical input are 
\be{} \lbrace (r^1,s^{1}),(r^2,s^{2}),(r^3,s^{3}),b_{1},b_2,\tb_{1},\tb_2\rbrace =\lbrace (-1,0),(0,-1),(0,-1),0,0,\frac{1}{2},\frac{1}{2}\rbrace. \non\ee
All the EOM are trivially invariant under $M^{(n)}_{l}$. The invariance under $L^{(n)}$ is given by 
\bea{}[L^{n},(\ref{113eq1})]=(n+1)(\Delta_1 -1)[nt^{n-1}\p_j A^{1}_{t}+gt^{n}\lbrace (\p^{i}A^{2}_{i})A^{3}_{j} +2A^{2}_{i}(\p^{i}A^{3}_{j})\non\\ -A^{2}_{i}(\p_{j}A^{3}_{i}) -(\p^{i}A^{3}_{i})A^{2}_{j}-2A^{3}_{i}(\p^{i}A^{2}_{j})+A^{3}_{i}(\p_{j}A^{2}_{i}) \rbrace],\eea
\bea{}~[L^{n},(\ref{113eq2})]=(n+1)(\Delta_1 -1)[-nt^{n-1}\p^{i}A^{2}_{i}+gt^{n}(2A^{3}_{i}\p^{i}A^{1}_{t}+(\p^{i}A^{3}_{i})A^{1}_{t})],\\
~[L^{n},(\ref{113eq3})]=-(n+1)(\Delta_1 -1)[nt^{n-1}\p^{i}A^{3}_{i}+gt^{n}(2A^{2}_{i}\p^{i}A^{1}_{t}+(\p^{i}A^{2}_{i})A^{1}_{t})],
\eea
\bea{}~[L^{n},(\ref{113eq4})]=in(n+1)t^{n-1}(\Delta_2 -\frac{3}{2})\phi_{1}+(\Delta_1 -1)t^n (n+1)(gT^{1}_{12}A^{1}_{t}\phi_2),\\
~[L^{n},(\ref{113eq5})]=in(n+1)t^{n-1}(\Delta_2 -\frac{3}{2})\phi_{2}+(\Delta_1 -1)t^n (n+1)(gT^{1}_{21}A^{1}_{t}\phi_1),
\eea
\be{}[L^{n},\p_i (\p^i A^{j(2,3)}-\p^j A^{i(2,3)})]=0,~[L^{n},\p_i\p^i A_{t}^{1}]=0,~[L^{n},i\sigma^{i}\p_{i}\phi_{1,2}]=0. \ee
We see that the EOM are invariant under $L_{n}$.
\subsubsection*{MMM case:}
The gauge fields and spinors scales as
 \be{} A_{i}^{a} \rightarrow  A_{i}^{a},~
A_{t}^{a} \rightarrow \epsilon A_{t}^{a},~\phi_{1,2} \rightarrow  \phi_{1,2} ,~\chi_{1,2} \rightarrow \epsilon\chi_{1,2}.\ee
EOM in this limit are given by
\bes
\bea{} \p_{i}(\p^i A^{ja}-\p^j A^{ia})=0,~
 \p_i (\p^i A_{t}^{a}-\p_t A^{ia})=0,\\i\sigma^{i}\p_{i}\phi_{1,2}=0,~ i\p_t\phi_{1,2}+i\sigma^{i}\p_{i}\chi_{1,2}=0.
\eea
\ees
Next, we shall look at the invariance of the equations under $L^{(n)}$'s. The values of the coefficients, as fixed by dynamics, are  
\be{} \lbrace r^a,s^{a},b_1,b_2,\tb_1,\tb_2\rbrace =\lbrace 0,-1,0,0,\frac{1}{2},\frac{1}{2}\rbrace. \non\ee
All the EOM are trivially invariant under $M^{(n)}_{l}$. The invariance under $L^{(n)}$ is given by 
\bea{} [L_n, \p_{i}(\p^i A^{ja}-\p^j A^{ia})]=0,~[L_n, i\sigma^{i}\p_{i}\phi_{1,2}]=0,\\
~[L_n, \p_i (\p^i A_{t}^{a}-\p_t A^{ia})] =-n(n+1)t^{n-1}(\Delta_1 -1)\p^i A^{a}_{i}, \\
~[L_n, i\p_t\phi_{1,2}+i\sigma^{i}\p_{i}\chi_{1,2}] = in(n+1)t^{n-1} (\Delta_2 -\frac{3}{2})\phi_{1,2}.
\eea
As seen above, the EOMs are invariant under $L_{n}$.
 
\section{Discussions}

In this paper, we have performed a detailed analysis of Galilean field theories and examined their symmetries. In particular, we have focussed on theories which are obtained by a systematic non-relativistic limit from a parent relativistic conformal field theory. We started out by considering free theories like the massless scalars and fermions and revisited our previous constructions of Galilean gauge theories, viz. Galilean electrodynamics and Yang-Mills theory. We then went on to consider matter (both scalars and fermions) added to these Galilean gauge theories. 

\subsection*{A conjecture} 

We saw that in all our field theory examples, if the parent theory exhibited relativistic conformal invariance, the theory obtained in the limit was invariant under the {\em{infinite-dimensional}} Galilean conformal algebra. 
We thus conjecture the following: 

\medskip

\noindent{{\em{\underline{Any} relativistic conformal field theory, in {\underline{any}} spacetime dimension, contains a non-relativistic subsector, the symmetries of which are dictated by the {\underline{infinite dimensional}} Galilean conformal algebra.}} 

\medskip

\noindent So, we are claiming that there is a generic {\em{infinite enhancement}} of symmetries in the non-relativistic limit of any CFT, even for spacetime dimensions $D>2$. We have shown that this is the case for a very wide variety of examples. Of course, a proof of this is still lacking and there are several unanswered questions. Let us list some of these and some related puzzles below. 

\subsection*{The $SU(2)$ puzzle}

Our first puzzle already manifests itself in our present work. As is explained in detail in the Appendix \ref{ap3}, if we take arbitrary scaling of fermions in the four sectors of the NR $SU(2)$ theory, there are around $\mathcal{O}(1500)$ different limits that one could construct. As we have already mentioned in the previous section, we consider restrictions on our limits of the following form: 
\begin{itemize}
\item The limit should give back the free NR Dirac equations \refb{eomnr} when the gauge fields are turned off.
\item The EOM in the limit should reduce to appropriate Galilean $SU(2)$ equations \refb{gymeom3}-\refb{gymeom2} when matter fields are turned off. 
\end{itemize}
Amazingly, these consistency requirements immediately bring down the $\mathcal{O}(1500)$ possibilities to 46 different limits in the $SU(2)$ theory. Again, rather surprisingly, these 46 sectors lead to only 19 different sets of EOM, after accounting for various exchange symmetries. If we wish to discard ``uninteresting" sectors where there are no fermion-gauge field interaction terms, we get a further reduction to a set of 15. As we have stated before, all these different sets of EOM exhibit invariance under the infinite GCA. This enormous reduction in the number of possible distinct sectors in the NR theory seems to indicate that there is something deeper at play which we are missing. It is also possible that by some more consistency requirements, we would be able to cut down further on the possible non-relativistic sectors of the $SU(2)$ theory. It is, of course, extremely important to understand the general structure of the limit in the $SU(2)$ theory before embarking on an analysis of the generic $SU(N)$ theory. 

\subsection*{Further remarks}
Recently, \cite{Festuccia:2016caf} re-examined the symmetry structure of non-relativistic electrodynamics and found that the EOM in both the electric and magnetic limit exhibited infinite dimensional symmetries. The symmetries discovered were even larger than we discussed in our earlier work \cite{Bagchi:2014ysa}, but contained the GCA as an infinite sub-algebra. It appears that the reason why the authors of \cite{Festuccia:2016caf} discovered more symmetries than our earlier work \cite{Bagchi:2014ysa} was that they did not consider that the values of $\Delta$ obtained from the relativistic Maxwell theory, and chose one which led to the enhanced symmetries. They also found that the infinite symmetries existed for all dimensions, which again boiled down to this freedom in the choice of $\Delta$. The curiosity of this observation is that usual Maxwell theory is only relativistic conformally invariant in $D=4$ and not in higher $D$. So it seems that by tinkering around with the non-relativistic conformal weights, one can perhaps eradicate anomalies in relativistic theories in their non-relativistic limit.  Also, the analysis of \cite{Festuccia:2016caf} would point to the fact that non-relativistic conformal systems perhaps exist even without the existence of any parent relativistic conformal theory, thus making the parameter space of non-relativistic conformal systems even richer that the relativistic ones. We would like to revisit our analysis of matter fields coupled to Galilean gauge theories in the framework of \cite{Festuccia:2016caf} to see whether more symmetries emerge by relaxing conditions on the Galilean conformal weights.  

As we just mentioned, the existence of extended infinite dimensional symmetries in the Galilean limit of Electrodynamics even for $D>4$ in \cite{Festuccia:2016caf} raises interesting possibilities about the anomaly structure in Galilean theories. In particular, it seems to suggest that $L_1$ and $M_1$, which are the descendants of the relativistic special conformal generator, curiously become symmetries although this was clearly not the case in the relativistic Maxwell theory. If there are no subtle loopholes in the arguments of \cite{Festuccia:2016caf}, there seems to be a very real possibility that the existence of infinite symmetries may wash away anomalies arising from parent relativistic theories. Some work on non-relativistic anomalies can be found in \cite{Jensen:2014hqa, Jain:2015jla}. We wish to re-examine Galilean anomalies from our limiting perspective. Explicitly, we wish to address the possible restoration of Galilean conformal symmetries in the non-relativistic version of Quantum Electrodynamics. 

A very natural avenue to which we would like to extend our work is Supersymmetric gauge theories. In particular, we would like to examine $\mathcal{N}=4$ Super Yang-MIlls (SYM) theory. This is one of the rare relativistic field theories that is conformally invariant in the quantum regime. It is expected that non-relativistic sectors of $\mathcal{N}=4$ SYM would exhibit the infinitely extended (appropriately supersymmetrised version of) GCA as its underlying symmetry algebra{\footnote{See e.g. \cite{deAzcarraga:2009ch, Sakaguchi:2009de, Bagchi:2009ke} for attempts at supersymmetrising the GCA in $D>2$.}}. This would be the first step to a whole host of questions, some of which we alluded to in the introduction. The emergence of infinite symmetries are pointers to integrability. If indeed the Galilean version of SYM exhibit infinite symmetries, this could indicate that these are sectors which exhibit quantum integrability and this would thus be a new integrable sector in SYM over and beyond the usual planar sector. Examining the planar sector in the limit should also be an interesting exercise as existing relativistic integrable sectors would perhaps get augmented by newer structures appearing from the symmetry enhancement in the limit. 

$\mathcal{N}=4$ SYM is of course dual to Type IIB string theory on AdS$_5 \times$S$^5$. So we should be able to explore the dual theory to the non-relativistic sectors of $\mathcal{N}=4$ SYM. Here we should be able to put on firm footing the idea of a Newton-Cartan like AdS$_2 \times \mathbb{R}^3$ emerging from the Galilean limit of AdS$_5$ as was put forward in \cite{Bagchi:2009my}. There are numerous possible directions of future work on the bulk side that would stem from a better understanding of this picture. 

On a similar note, Newton-Cartan structures would also emerge on the field theory side. It would be useful to formulate the field theoretic considerations in this and earlier works in a more geometric formalism that would be intimately linked to Newton-Cartan geometry. This should also facilitate an action formulation of these theories which we currently lack{\footnote{See however \cite{Bergshoeff:2015sic, Festuccia:2016caf} for some action formulations that necessitate the incorporation of auxiliary fields.}}.

\subsection*{Carrollian field theories} 

Finally, we would like to comment on a related field where our present considerations would be very useful. Instead of taking the speed of light to infinity, one can consider rather peculiar theories where the speed of light is taken to zero. Field theories with this feature are called Carrollian field theories and are closely related to their Galilean cousins. Carrollian field theories have recently emerged as theories of interest as their conformal extensions, Carrollian CFTs have been found to be putative dual theories to Minkowskian spacetimes. The Carrollian Conformal Algebra (CCA) is isomorphic to the Bondi-Metzner-Sachs (BMS) algebra in one higher dimension. The BMS algebra is the asymptotic symmetry algebra on the null boundary of Minkowski spacetimes \cite{Bondi:1962px, Sachs:1962zza} and it has been known since 1970's that these algebras in three and four dimensional Minkowski spacetimes are infinite dimensional. Recently, there has been a resurgence of activities in the investigation of infra-red physics related to the BMS group and there have been developments linking the BMS group to long known soft theorems and memory effects in an infrared triangle of relations. For a review of these topics, the reader is referred to \cite{Strominger:2017zoo} and the references within it. 

In three bulk spacetime dimensions, the BMS algebra takes the form \cite{Barnich:2006av}
\begin{subequations}\label{bms3}
\bea{}   
&& [L_n, L_m] = (n-m) L_{n+m} + \frac{c_L}{12} \delta_{n+m, 0} (n^3 - n) \\
&& [L_n, M_m] = (n-m) M_{n+m} + \frac{c_M}{12} \delta_{n+m, 0} (n^3 - n)\\
&& [M_n, M_m] = 0
\eea 
\end{subequations}
Here $L_n$ are the so-called super-rotations, that form the Diff$(S^1)$ of the circle at null infinity and the $M_n$ are the super-translations, which are angle dependent translations of the null direction. The central terms $c_L$ and $c_M$ arise in the algebra and for Einstein gravity \cite{Barnich:2006av}, they are $c_L=0, \ c_M=1/4G$, where $G$ is the Newton's constant. The attentive reader would have noticed already that this above algebra \refb{bms3} is isomorphic to the GCA in 2d which follows from \refb{GCA}. This connection was noticed first in \cite{Bagchi:2010eg} and later exploited in a variety of works (e.g. \cite{Bagchi:2012yk} -- \cite{Barnich:2012rz}) in an attempt to formulate the holographic correspondence for 3d flat spacetimes. The reader is referred to \cite{Bagchi:2016bcd, Riegler:2016hah} for a more comprehensive review of the field. The use of the 2d GCA in this case is justified as the contractions $c\to \infty$ and $c\to 0$ yields the same algebras starting from two copies of the Virasoro algebra. In other words, the GCA and CCA are isomorphic in 2d. Interestingly, this algebra has also appeared as symmetries of the worldsheet of tensionless strings \cite{Bagchi:2013bga,Bagchi:2015nca} and in the context of the ambi-twistor string \cite{Casali:2016atr}.  

In higher dimensions, the GCA and CCA differ as the number of contracted directions for the GCA and the CCA don't remain the same. The GCA is a non-relativistic limit and hence, as we have discussed through out the present paper, the contraction to use is $x_i \to \e x_i, \, t \to t$. The CCA, on the other hand, requires us to take the speed of light to zero and hence the contraction required is $x_i \to x_i, \, t \to \e t$. In \cite{Bagchi:2016bcd} , gauge theories in $D=4$, {\it{viz.}} Electrodynamics (see also \cite{Duval:2014uoa}) and Yang-Mills theories, in the Carrollian limit were constructed and it was found that infinite dimensional enhancements similar in spirit (but different in the actual algebraic details) to the ones we have discussed in this paper are also present there. An on-going investigation is whether matter added to these Carrollian field theories would also generate the infinite conformal structures, now in the ultra-relativistic regime. These considerations would be very useful for understanding the flat limit of the parent Maldacena correspondence that relates AdS$_5 \times$ S$^5$ to $\mathcal{N}=4$ SYM. The comments about integrability and anomalies that we mentioned earlier would also be very pertinent in the ultra-relativistic context. 

\bigskip

\section*{Acknowledgements}
It is a pleasure to thank Rudranil Basu, Poulami Nandi, Kostas Skenderis, Wei Song and Marika Taylor for helpful discussions and Sunando Patra for help with {\it{Mathematica}} for the detailed calculations of the appendices. 

\medskip

\noindent We would like to acknowledge the hospitality of the following institutes/universities during various stages of this work: Albert Einstein Institute (AM), University of Southampton (AB), Universite Libre de Bruxelles (AB, AM), Vienna University of Technology (AB, AM), Tsinghua University, Beijing (AB). The work is partially supported by DST Inspire faculty fellowships (AB and JC), a Max Planck mobility award (AB), and an SERB Early Career Research Award (JC). 

\bigskip \bigskip

\appendix
\section*{APPENDICES}
\section{Conditions of scaling of Scalar fields}
Here, to start with, we induce most general  scaling of the scalar fields ($\alpha~\text{and}~\beta$) along with the scaling of gauge fields.
Then we note down the inequalities between $\alpha$ and $\beta$ which are consistent with EOM of free theory. With the allowed values of  $\alpha$ and  $\beta$ we  compute the non-relativistic equation of motions in the limit $\epsilon \to 0$. It is worthy to mention that only  a particular choice of $\alpha$ and  $\beta$  leaves some relics of the interacting theory.
The relativistic EOM for scalar electrodynamics are given as 
\bea{} \p^{\mu}F_{\mu\nu}+e[\psi_{1}\p_{\nu}\psi_{2}- \psi_{2}\p_{\nu}\psi_{1}-eA_{\nu}(\psi^{2}_{1}+\psi^{2}_{2})]=0,\\
\p^{\mu}\p_{\mu}\psi_1 + e\p^{\mu}A_{\mu}\psi_2 +2eA_{\mu}\p^{\mu}\psi_2 -e^2 A^{\mu}A_{\mu}\psi_1=0, \\
\p^{\mu}\p_{\mu}\psi_2 - e\p^{\mu}A_{\mu}\psi_1 -2eA_{\mu}\p^{\mu}\psi_1 -e^2 A^{\mu}A_{\mu}\psi_2=0. \eea  
Below we use the electric and magnetic limits of  gauge fields to find out the interplay between $\alpha$ and $\beta$ to match up our requirement.
\subsubsection*{Electric sector} 
The gauge and the scalar fields are scaled in the electric limit  as
\be{} A_{t} \rightarrow A_t ,~~ A_i \rightarrow \epsilon A_{i},~~ \psi_1 \rightarrow \epsilon^{\alpha} \psi_1 ,~~\psi_2 \rightarrow \epsilon^{\beta} \psi_2,  \ee
and the EOM are read as
\bea{} \p^i \p_i A_t -\epsilon^2 \p^i \p_t A_i +e[\epsilon^{\alpha +\beta+2}(\psi_1 \p_t \psi_2 - \psi_2 \p_t \psi_1)-eA_t (\epsilon^{2\alpha+2}\psi^{2}_{1}+\epsilon^{2\beta +2}\psi^{2}_{2})]=0,\eea
\bea{} (\p^i \p_i A_j -\p^i \p_j A_i +\p_t \p_j A_t) - \epsilon^{2}\p_t \p_t A_j +e[\epsilon^{\alpha +\beta}(\psi_1 \p_j \psi_2 - \psi_2 \p_j \psi_1) \non\\- eA_j (\epsilon^{2\alpha+2}\psi^{2}_{1}+\epsilon^{2\beta+2}\psi^{2}_{2})]=0.\eea
We find the following relations
\be{} \alpha+\beta+2 \geqslant 0,~\alpha+1\geqslant 0,~ \beta+1\geqslant 0,~\alpha+\beta\geqslant0, \ee
that are compatible with EOM of free theory. The allowed region in $\alpha-\beta$ plane is shown in Fig.~\ref{fig:scalar_E}. The interaction terms disappear for all other choice except $\alpha=1,\beta=-1$. The non-relativistic equations for this particular set of values are given in (\ref{esed1}).

\subsubsection*{Magnetic sector} 
The scaling of  gauge  and  scalar fields are given in magnetic limit as
\be{} A_{t} \rightarrow\epsilon A_t ,~~ A_i \rightarrow  A_{i},~~ \psi_1 \rightarrow \epsilon^{\alpha} \psi_1 ,~~\psi_2 \rightarrow \epsilon^{\beta} \psi_2, \ee
and the EOM are read as
\bea{} (\p^i \p_i A_t - \p^i \p_t A_i) +e[\epsilon^{\alpha +\beta+1}(\psi_1 \p_t \psi_2 - \psi_2 \p_t \psi_1)-eA_t (\epsilon^{2\alpha+2}\psi^{2}_{1}+\epsilon^{2\beta +2}\psi^{2}_{2})]=0,\eea
\bea{} (\p^i \p_i A_j -\p^i \p_j A_i) -\epsilon^2 ( \p_t \p_t A_j -\p_t \p_j A_t) +e[\epsilon^{\alpha +\beta+1}(\psi_1 \p_j \psi_2 - \psi_2 \p_j \psi_1) \non\\- eA_j (\epsilon^{2\alpha+2}\psi^{2}_{1}+\epsilon^{2\beta+2}\psi^{2}_{2})]=0.\eea
We derive the inequalities that satisfy the EOM of the free theory as
\be{} \alpha+\beta+1 \geqslant 0,~\alpha+1\geqslant 0,~ \beta+1\geqslant 0.\ee
The allowed domain is depicted in Fig.~\ref{fig:scalar_M}.
The interactions are non vanishing only with  $\alpha=0 , \beta =-1$, and the derived non-relativistic equations are given in  (\ref{msed1}).
\begin{figure}[h!]
	\centering
	\subfloat[Electric case]{
		\includegraphics[scale=0.45]{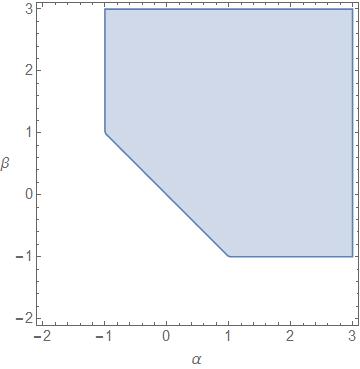}
		\label{fig:scalar_E}}
	\subfloat[Magnetic case]{
		\includegraphics[scale=0.45]{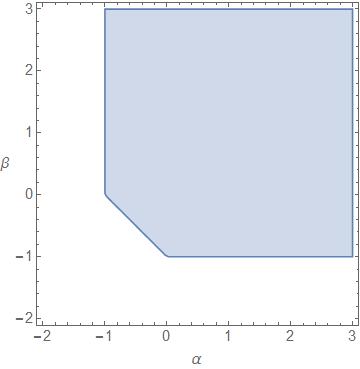}
		\label{fig:scalar_M}}
	\caption{Conditions for scaling of scalar fields in Electric and Magnetic limits.}
	\label{figu}
\end{figure}

\section{Conditions of scaling of Fermionic fields}\label{ApB}
We look at the arbitrary scaling of fermionic fields in the free and interacting theories. We then try to find the conditions on the scaling by plugging them into the relativistic EOM. If the conditions are satisfied, we get back the free or interacting non-relativistic EOM. \\We also focus on how to incorporate the non-relativistic scaling of the fermions in the Weyl representation. By doing so, we want to show that the non-relativistic equations we got are representation independent.  
\subsection{Free theory}
We start with the massless Dirac equation
\be{} i\gamma^{\mu}\p_{\mu}\Psi=0. \ee
Using Pauli-Dirac representation (\ref{compbr}) and (\ref{gama1}), the EOM become
\be{}\label{eom1} i\p_t \phi + i\sigma^i \p_i \chi =0 ,~~  i\p_t \chi + i\sigma^i \p_i \phi =0.
\ee
Now we consider the following scaling of the fermion fields
\be{} \phi \rightarrow \epsilon^{\alpha}\phi,~ \chi \rightarrow \epsilon^{\beta}\chi\ee
which are plugged  into (\ref{eom1}), and we find
\be{} i\p_t \phi +\epsilon^{\beta-\alpha-1} i\sigma^i \p_i \chi =0 ,~~ i\p_t \chi +  \epsilon^{\alpha-\beta -1} i\sigma^i \p_i \phi =0.
\ee
We note  the constraint for these scalings  are given as 
\be{impcon} \beta-\alpha = 1 ~\text{or}~ \alpha -\beta =1. \ee
The non-relativistic equations of massless fermion theory
\be{}\label{nreom} i\p_t \phi + i\sigma^i \p_i \chi =0,~  i\sigma^i \p_i \phi =0. \ee
are achieved for  $\alpha=0, \beta=1$.

The relation between the Weyl representation and Pauli Dirac representation is given by:
\bea{} \gamma^{\mu}_{W}= S\gamma^{\mu}_{D}S^{-1} , \Psi_W =
\begin{pmatrix}
u \\
v \\
\end{pmatrix}
=S\Psi_D =S\begin{pmatrix}
\phi \\
\chi \\
\end{pmatrix},
\eea
where `W' and `D' stand for Weyl and Pauli-Dirac representations respectively with $S =\frac{1}{\sqrt{2}}(1+\gamma_5 \gamma^0)$. The gamma matrices are in the Pauli-Dirac representation and using the scaling of fields, i.e., $\phi \rightarrow \phi,~ \chi \rightarrow \epsilon \chi$, we find
\be{}\label{we0} u = \frac{1}{\sqrt{2}} (\phi - \epsilon\chi),~ v = \frac{1}{\sqrt{2}} (\phi + \epsilon\chi). \ee
Feeding (\ref{we0}) into the Weyl equations $(i\sigma^{\mu}\p_{\mu}v=0 ,~ i\bar{\sigma}^{\mu}\p_{\mu}u =0)$, we find
\be{} i\p_t \phi + i\sigma^i \p_i \chi =0 ,~~  i\sigma^i \p_i \phi =0.
\ee
The EOM in the non-relativistic limit are thus representation independent. In passing we note that we can choose $\alpha=1,\beta=0$ and this leads to the same equations as (\ref{nreom}) with $\phi \leftrightarrow \chi$.

\subsection{Galilean Yukawa theory}
We consider a Yukawa  theory with an underlying $U(1)$ global symmetry. Here,  $\varphi$ and $\Psi$ are the singlet scalar, and  non-singlet Dirac fields respectively with  coupling constant $g$.
The relativistic EOM are
\be{} \p^{\mu}\p_{\mu}\varphi -g(\phi^{\dagger}\phi -\chi^{\dagger}\chi)=0,~
i\p_{t}\phi +i\sigma^{i}\p_{i}\chi -g\varphi\phi=0,~
i\p_{t}\chi +i\sigma^{i}\p_{i}\phi +g\varphi\chi=0.\ee
Consider the scaling of the scalar ($\varphi$)  and fermion ($\phi,\chi$) fields as
\be{} \varphi\rightarrow \varphi,~ \phi \rightarrow \epsilon^{\a}\phi ,~ \chi \rightarrow \epsilon^{\b} \chi. \ee
In this limit, the EOM become
\bea{} -\epsilon^2 \p_{t}\p_{t}\varphi + \p^{i}\p_{i}\varphi - g(\epsilon^{2\a+2}\phi^{\dagger}\phi -\epsilon^{2\b+2}\chi^{\dagger}\chi)=0,\\
i\p_{t}\phi +\epsilon^{\b-1-\a}i\sigma^{i}\p_{i}\chi -g\varphi\phi=0,\\
i\p_{t}\chi +\epsilon^{\a-1-\b}i\sigma^{i}\p_{i}\phi +g\varphi\chi=0,
\eea
leading to the following relation among  the scaling of fermions 
\be{} \a+1\geq0,\b+1\geq0, \beta-\alpha-1= 0 ~\text{or}~ \alpha -\beta-1 =0. \ee
The non-relativistic equations of massless Yukawa theory considered in the main text are regained with  $\alpha=0, \beta=1$.
\subsection{Galilean Electrodynamics with Fermions}
The relativistic EOMs for spinor Electrodynamics  are given as
\bea{} \p^{i}F_{it}-e(\phi^{\dagger}\phi+\chi^{\dagger}\chi)=0,~
-\p_{t}F_{tj}+\p^{i}F_{ij}+e(\phi^{\dagger}\sigma^{j}\chi
+\chi^{\dagger}\sigma^{j}\phi)=0,\\
 (i\p_{t}+eA_{t})\phi +(i\sigma^{i}\p_{i}+e\sigma^{i}A_{i})\chi=0 ,~
(i\sigma^{i}\p_{i}+e\sigma^{i}A_{i})\phi+ (i\p_{t}+eA_{t})\chi=0,
\eea
Here, we are interested in the scaling of spinor fields that are consistent with the EOM of free theory and also leaves some interaction terms between the fermion and gauge fields. 
\subsubsection*{Electric sector} 
 The scaling of the gauge fields and the scalar fields are given as
\be{} A_{t} \rightarrow A_t ,~~ A_i \rightarrow \epsilon A_{i},~~ \phi \rightarrow \epsilon^{\alpha} \phi ,~~\chi \rightarrow \epsilon^{\beta} \chi,  \ee
and the EOM are
\bes \label{eomesfm}
\bea{} \p^i \p_i A_t -\epsilon^2 \p^i \p_t A_i -e[\epsilon^{2\alpha+2}\phi^{\dagger}\phi+\epsilon^{2\beta +2}\chi^{\dagger}\chi]=0,\eea
\bea{} (\p^i \p_i A^j -\p_i \p^j A^i +\p_t \p^j A_t) - \epsilon^{2}\p_t \p_t A^j +e[\epsilon^{\alpha+\beta+1}(\phi^{\dagger}\sigma^j \chi + \chi^{\dagger}\sigma^j \phi)]=0.\eea\ees
The following relations are achieved to satisfy our criteria:
\be{} \alpha+\beta+1 \geqslant 0,~\alpha+1\geqslant 0,~ \beta+1\geqslant 0 \ee
along with (\ref{impcon}). The equations (\ref{esse}) are achieved with $\alpha=-1 , \beta =0$. If we take the values as $\alpha=0 , \beta =1$ and plug it into (\ref{eomesfm}) and in the fermionic equations, we again get back the same interaction pieces given in (\ref{eomesf}). We note that  these constants possess two set of values which correspond to two sets of equations with same interaction pieces.    

\begin{figure}[h!]
	\centering
	\subfloat[Electric case]{
		\includegraphics[scale=0.45]{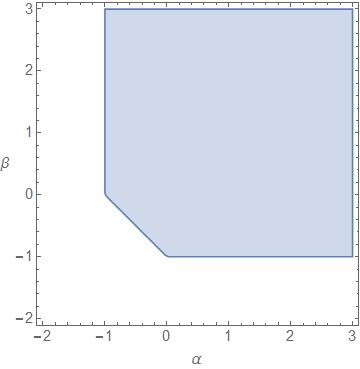}
		\label{fig:224-dim5-dodd_1}}
	\subfloat[Magnetic case]{
		\includegraphics[scale=0.45]{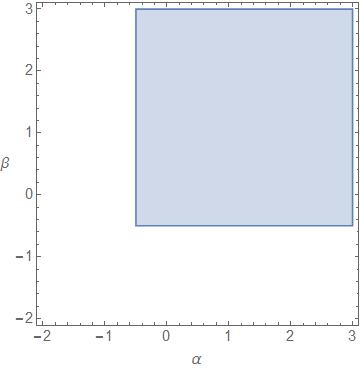}
		\label{fig:224-dim5-dodd_2}}
	\caption{Conditions for scaling of spinor fields in Electric and Magnetic limits.}
	\label{fig:224-dim5-dodd}
\end{figure}

\subsubsection*{Magnetic sector} 
 In this limit, the fields go through the following scaling:
\be{} A_{t} \rightarrow\epsilon A_t ,~~ A_i \rightarrow  A_{i},~~ \phi \rightarrow \epsilon^{\alpha} \phi ,~~\chi \rightarrow \epsilon^{\beta} \chi,  \ee
and the EOMs are read as 
\bea{} (\p^i \p_i A_t - \p^i \p_t A_i) -e[\epsilon^{2\alpha+1}\phi^{\dagger}\phi+\epsilon^{2\beta +1}\chi^{\dagger}\chi]=0,\eea
\bea{} (\p^i \p_i A^j -\p^i \p^j A_i) -\epsilon^2 (\p_t \p_t A^j -\p_t \p^j A_t ) +e[\epsilon^{\alpha +\beta+2}(\phi^{\dagger}\sigma^j \chi + \chi^{\dagger}\sigma^j \phi)]=0.\eea
Our demands are fulfilled   with
\be{} 2\alpha+1 \geqslant 0,~\alpha+\beta+2\geqslant 0,~ 2\beta+1\geqslant 0,\ee
along with (\ref{impcon}). The non-relativistic equations (\ref{msse}) are consistent with $\alpha=1 , \beta =0$. Similarly, for magnetic case, we can also take the values of the constants as $\alpha=-\frac{1}{2} , \beta =\frac{1}{2}$ and get back the same equations as (\ref{msse}) expect for one which is given as
\bea{} (\p^i \p_i A_t - \p^i \p_t A_i) -e\; \phi^{\dagger}\phi=0.\eea 
Due to this equation, the two different sets give rise to two different EOM.
We have to keep in mind about these limits of $U(1)$ case when we speak about the equations in $SU(2)$ case. It can be seen clearly that if we switch off the self interactions and decouple the fermions, we get back the $U(1)$ equations. 

\section{Scaling of spinors in $SU(2)$ Yang-Mills coupled with Fermions}\label{ap3}

To start with, we adopt the most general scaling of fermionic fields within the $SU(2)$ Yang-Mills theory and  find out the consistent relations among these scalings. Then we make a specific choice of  $\alpha~\text{and}~\beta$ and take the limit $\epsilon \rightarrow 0$ to get back the non-relativistic equations. 

Here, we consider the different sectors of Galilean $SU(2)$ Yang-Mills theory where fermions are in fundamental representation.  Then we implement  the electric and magnetic limits of  gauge fields to find out the interplay between $\alpha$ and $\beta$ to match up our requirement.
\subsection*{EEE sector}
In EEE sector of Galilean $SU(2)$ spinor Yang-Mills, the scaling of the fields are given as
\bea{scel} A_{i}^a \rightarrow \epsilon A_i^a ,~ A_t^a \rightarrow A_t^a,~\phi_1 \rightarrow \epsilon^{\a_1}\phi_1, \phi_2 \rightarrow \epsilon^{\a_2}\phi_2,~\chi_1 \rightarrow \epsilon^{\b_1}\chi_1,~\chi_2 \rightarrow \epsilon^{\b_2}\chi_2. \eea
Next, we  write the conditions of the scaling of spinors which we can found by using the fact that the EOM at least contains the free parts. The conditions are as follows:
\be{condeee}\alpha_{i}+\beta_{j}\geq -1 ,~\sum_{i}\alpha_{i}\geq -2,~\sum_{i}\beta_{i}\geq -2,~\alpha_{i}\geq -1,~\beta_{i}\geq -1 ~\text{with}~i,j=1,2.\ee
\begin{table}[htb]
\centering
\label{tabel} 
\begin{tabular}{|p{1cm}|p{1cm}|p{1cm}|p{1cm}|p{1cm}|p{5cm}|}
 \hline
 \multicolumn{6}{|c|}{EEE sector} \\
 \hline
 &$\a_1$& $\a_2$ &$\b_1$ & $\b_2$& EOM\\
 \hline
 1.&-1&-1&~0&~0&$A_e$\\
 \hline
 2.&-$\frac{1}{2}$& -$\frac{1}{2}$ &~$\frac{1}{2}$& ~$\frac{1}{2}$& $C_e$\\
  \hline
 3.&~0& ~0  & ~1&~1& $C_e$\\
 \hline 
 \hline
 4.&-1&~1&~0&~0&$B_e$\\
  \hline
 5.&-$\frac{1}{2}$& ~$\frac{1}{2}$  & ~$\frac{1}{2}$   &-$\frac{1}{2}$& $D_e$\\
  \hline
6.&-$\frac{1}{2}$ &~1 & ~$\frac{1}{2}$&~0& $E_e$\\
 \hline
 7.&~0& ~0  & ~1&-1& $B_e$ ($\phi_1 \Leftrightarrow \chi_2, \phi_2\Leftrightarrow\chi_1$)\\
 \hline
8.&~0& ~$\frac{1}{2}$  & ~1&-$\frac{1}{2}$& $E_e$\\
 \hline
9.&~0& ~1 & ~1&~0& $E_e$\\
\hline
\hline
10.&~0& ~0  & -1&-1& $A_e$ ($\phi_1\Leftrightarrow\chi_1,\phi_2\Leftrightarrow\chi_2$)\\
 \hline
 11.&~$\frac{1}{2}$&~$\frac{1}{2}$&-$\frac{1}{2}$&-$\frac{1}{2}$& $C_e$ ($\phi_1\Leftrightarrow\chi_1,\phi_2\Leftrightarrow\chi_2$)\\
   \hline
12.&~1& ~1  & ~0&~0& $C_e$ ($\phi_1\Leftrightarrow\chi_1,\phi_2\Leftrightarrow\chi_2$)\\
\hline
\hline
13.&~0&~0  & -1&~1& $B_e$ ($\phi_1\Leftrightarrow\chi_1,\phi_2\Leftrightarrow\chi_2$)\\
 \hline
14.&~$\frac{1}{2}$    &-$\frac{1}{2}$ & -$\frac{1}{2}$& ~$\frac{1}{2}$&$D_e$ ($\phi_1\Leftrightarrow\chi_1,\phi_2\Leftrightarrow\chi_2$)\\
 \hline
 15.&~$\frac{1}{2}$&~0 &-$\frac{1}{2}$ &~1&$E_e$ ($\phi_1\Leftrightarrow\chi_1,\phi_2\Leftrightarrow\chi_2$)\\
 \hline
16.&~1& -1  & ~0&~0& $B_e$ ($\phi_1\Leftrightarrow\phi_2,\chi_1\Leftrightarrow\chi_2$)\\
 \hline
17.&~1& -$\frac{1}{2}$  & ~0&~$\frac{1}{2}$&$E_e$ ($\phi_1\Leftrightarrow\chi_1,\phi_2\Leftrightarrow\chi_2$)\\
 \hline
18.&~1& ~0 & ~0&~1&$E_e$ ($\phi_1\Leftrightarrow\chi_1,\phi_2\Leftrightarrow\chi_2$)\\
 \hline
\end{tabular}
\caption{Scaling of spinors in EEE sector:  [1-3] satisfy $\b_{1,2}-\a_{1,2}=1$;~[4-9] satisfy $\b_{1}-\a_{1}=1$ and $\a_{2}-\b_{2}=1$;~[10-12] satisfy $\a_{1,2}-\b_{1,2}=1$;~[13-18] satisfy $\a_{1}-\b_{1}=1$ and $\b_{2}-\a_{2}=1$ along with (\ref{condeee}).}
\end {table}
In the table, we mention the cases where the structure of EOM remain invariant if we exchange fermion fields suitably.  For example, we can see that if we take the values of the coefficients ($\a_1=-1,\a_2=-1,\b_1=0,\b_2=0$), the corresponding EOM is denoted by $A_e$. Now, if we look at the case where ($\a_1=0,\a_2=0,\b_1=-1,\b_2=-1$), we see that the EOM is also denoted by $A_e$ but with the exchange symmetry given by ($\phi_1\Leftrightarrow\chi_1,\phi_2\Leftrightarrow\chi_2$).\\ To make our analysis bit simple, we will define certain quantities like $A_e,B_e,...$ that will be related to different sets of EOM. Since, we are in $SU(2)$ case, we will take $a=1,2,3$ and $m=1,2$.
The EOM are given as follows:\\
$\bullet{~\underline{A_{e} ~case}}:$
\bea{}
1) &&\p_t\p^j A_{t}^{a}+\p^{i}(\p_i A^{ja}-\p^j A_{i}^{a})+gf^{abc}A^{b}_{t}\p^{j}A_{t}^{c}+g(\phi^{\dagger}_{m} \sigma^{j}T^{a}_{mn}\chi_{n}
+\chi^{\dagger}_{m} \sigma^{j}T^{a}_{mn}\phi_{n})=0,\non\\2) 
&& \p^i\p_i A_{t}^{a}-g(\phi^{\dagger}_m T^{a}_{mn}\phi_{n})=0,~i\sigma^{i}\p_{i}\phi_{1,2}=0,\non\\3) && i\p_t\phi_{1} +gT^{a}_{11}A^{a}_{t}\phi_1 +gT^{a}_{12}A^{a}_{t}\phi_2 +i\sigma^{i}\p_{i}\chi_{1}=0,\non\\4)&& i\p_t\phi_2 +gT^{a}_{21}A^{a}_{t}\phi_1 +gT^{a}_{22}A^{a}_{t}\phi_2 +i\sigma^{i}\p_{i}\chi_{2}=0. \non
\eea
$\bullet{~\underline{B_{e} ~case}}:$
\bea{}1) && \p_t\p^j A_{t}^{a}+\p^{i}(\p_i A^{ja}-\p^j A_{i}^{a})+gf^{abc}A^{b}_{t}\p^{j}A_{t}^{c}+g(\phi^{\dagger}_{1} \sigma^{j}T^{a}_{1n}\chi_{n}
+\chi^{\dagger}_{n} \sigma^{j}T^{a}_{n1}\phi_{1})=0,\non\\ 2)&&
\p_i\p^i A_{t}^{a}-g(\phi^{\dagger}_1 T^{a}_{11}\phi_{1})=0,~i\sigma^{i}\p_{i}\phi_{1}=0,\non\\3)&& gT^{a}_{21}A^{a}_{t}\phi_1 +i\sigma^{i}\p_{i}\chi_{2}=0,~ i\p_t\phi_{1} +gT^{a}_{11}A^{a}_{t}\phi_1 +i\sigma^{i}\p_{i}\chi_{1}=0,\non\\4)&& i\p_t\chi_2 +gT^{a}_{21}A^{a}_{t}\chi_1 +gT^{a}_{22}A^{a}_{t}\chi_2 +i\sigma^{i}\p_{i}\phi_{2}+g\sigma^{i}T^{a}_{21}A^{a}_{i}\phi_1 =0. \non
\eea
$\bullet{~\underline{C_{e} ~case}}:$
\bea{}
1)&&\p_t\p^j A_{t}^{a}+\p^{i}(\p_i A^{ja}-\p^j A_{i}^{a})+gf^{abc}A^{b}_{t}\p^{j}A_{t}^{c}=0,~
\p_i\p^i A_{t}^{a}=0,~ i\sigma^{i}\p_{i}\phi_{1,2}=0,\non\\2)&& i\p_t\phi_{1} +gT^{a}_{11}A^{a}_{t}\phi_1 +gT^{a}_{12}A^{a}_{t}\phi_2 +i\sigma^{i}\p_{i}\chi_{1}=0,\non\\3)&&   i\p_t\phi_{2} +gT^{a}_{21}A^{a}_{t}\phi_1 +gT^{a}_{22}A^{a}_{t}\phi_2 +i\sigma^{i}\p_{i}\chi_{2}=0. \non
\eea
$\bullet{~\underline{D_{e} ~case}}:$
\bea{}1)&& \p_t\p^j A_{t}^{a}+\p^{i}(\p_i A^{ja}-\p^j A_{i}^{a})+gf^{abc}A^{b}_{t}\p^{j}A_{t}^{c}+g(\phi^{\dagger}_{1} \sigma^{j}T^{a}_{12}\chi_{2}
+\chi^{\dagger}_{2} \sigma^{j}T^{a}_{21}\phi_{1})=0,\non\\2)&&
\p_i\p^i A_{t}^{a}=0,~i\sigma^{i}\p_{i}\phi_{1}=0, i\sigma^{i}\p_{i}\chi_{2}=0,\non\\3)&& i\p_t\phi_{1} +gT^{a}_{11}A^{a}_{t}\phi_1 +i\sigma^{i}\p_{i}\chi_{1}=0,\non\\4)&& i\p_t\chi_2 +gT^{a}_{22}A^{a}_{t}\chi_2 +i\sigma^{i}\p_{i}\phi_{2} =0. \non
\eea
$\bullet{~\underline{E_{e} ~case}}:$
\bea{}1)&&\p_t\p^j A_{t}^{a}+\p^{i}(\p_i A^{ja}-\p^j A_{i}^{a})+gf^{abc}A^{b}_{t}\p^{j}A_{t}^{c}=0,~
\p_i\p^i A_{t}^{a}=0,~i\sigma^{i}\p_{i}\phi_{1}=0\non\\2)&& i\sigma^{i}\p_{i}\chi_{2}=0,~ i\p_t\phi_{1} +gT^{a}_{11}A^{a}_{t}\phi_1 +i\sigma^{i}\p_{i}\chi_{1}=0,\non\\3)&& i\p_t\chi_2 +gT^{a}_{22}A^{a}_{t}\chi_2 +i\sigma^{i}\p_{i}\phi_{2} =0. \non
\eea

\subsubsection*{Gauge Transformation:}
We now look at the gauge transformation in this limit. The relativistic Yang-Mills theory is invariant under gauge transformations of the form 
\bea{}\label{gsym123}
\delta\psi_m= i\theta^a (T^{a}\psi)_m , ~~~ \delta A_{\mu}^a =\frac{1}{g}\p_{\mu}\th^a +f^{abc}A^{b}_{\mu}\th^c, ~~~ \eea
where $f^{abc}$ are the structure constant. The gauge transformation of $(\phi,\chi)$ fields are given as
\be{}\delta\phi_m= i\th^a T^{a}_{ml}\phi_l ,~~\delta\chi_m= i\th^a T^{a}_{ml}\chi_l.\ee
We will generalize our strategy that we have done in details in \cite{Bagchi:2015qcw}. We apply the scaling (\ref{scel}) on equation (\ref{gsym123}) and check what scaling of $\th^a$ keeps the
equation finite. We find that the scaling needs to be 
\be{} \th^{a} \rightarrow \epsilon^2 \th^{a}.\ee
The gauge transformation in this limit are
\be{elgl}\delta A^{a}_{t}=0,~\delta A^{a}_{i}=\frac{1}{g}\p_i \th^{a}.\ee
Similarly, for $(\phi,\chi)$ fields
\bea{spsl}\non \delta\phi_1 =\epsilon^2 (i\th^a T^{a}_{11}\phi_1)+\epsilon^{\a_2 -\a_1 +2} (i\theta^a T^{a}_{12}\phi_2),~ \delta\phi_2 =\epsilon^{\a_1 -\a_2 +2} (i\th^a T^{a}_{21}\phi_1)+\epsilon^{2} (i\th^a T^{a}_{22}\phi_2),\\ \delta\chi_1 =\epsilon^2 (i\th^a T^{a}_{11}\chi_1)+\epsilon^{\b_2 -\b_1 +2} (i\th^a T^{a}_{12}\chi_2),~ \delta\chi_2 =\epsilon^{\b_1 -\b_2 +2} (i\th^a T^{a}_{21}\chi_1)+\epsilon^{2} (i\th^a T^{a}_{22}\chi_2).
\cr
\eea 
Taking one set of values of the coefficients $(\a_1,\a_2,\b_1,\b_2)$ from the table along with $\epsilon\rightarrow 0$ limit and plugging them into (\ref{spsl}) gives us the required gauge transformation for the spinors in that sub case of the EEE sector. Together with (\ref{elgl}), we can then find the gauge invariance of the EOM associated with the given values of the coefficients.   
\subsection*{EEM sector}
The non-relativistic scaling of the fields in EEM sector are given as
\bea{} A_{i}^{1,2} \rightarrow \epsilon A_i^{1,2} ,~ A_t^{1,2} \rightarrow A_t^{1,2},~ A_{i}^{3} \rightarrow A_i^{3} ,~ A_t^{3} \rightarrow \epsilon  A_t^{3},\non\\ \phi_1 \rightarrow \epsilon^{\a_1}\phi_1 , \phi_2 \rightarrow \epsilon^{\a_2}\phi_2,~\chi_1 \rightarrow \epsilon^{\b_1}\chi_1,~\chi_2 \rightarrow \epsilon^{\b_2}\chi_2. \eea
The conditions of the scaling of spinors are 
\be{condeem}\alpha_{i}+\beta_{j}\geq -1 ,~\sum_{i}\alpha_{i}\geq -1,~\sum_{i}\beta_{i}\geq -1,~\alpha_{i}\geq -\frac{1}{2},~\beta_{i}\geq -\frac{1}{2} ~\text{with}~i,j=1,2.\ee
We  define the quantities like $A_{eem},B_{eem},...$ that will be related to different sets of EOM.
\begin{table}[htb]
\centering
\label{tabml} 
\begin{tabular}{|p{1cm}|p{1cm}|p{1cm}|p{1cm}|p{1cm}|p{5cm}|}
 \hline
 \multicolumn{6}{|c|}{EEM sector} \\
 \hline
 &$\a_1$& $\a_2$ &$\b_1$ & $\b_2$& EOM\\
 \hline
 1.&-$\frac{1}{2}$& -$\frac{1}{2}$ &~$\frac{1}{2}$& ~$\frac{1}{2}$&$A_{eem}$\\
  \hline
  2.&~0& ~0  & ~1&~1&$C_{eem}$\\
 \hline
 \hline
 3.&-$\frac{1}{2}$& ~$\frac{1}{2}$  & ~$\frac{1}{2}$   &-$\frac{1}{2}$&$B_{eem}$\\
 \hline
4.&~0& ~1 & ~1&~0&$F_{eem}$\\
 \hline
 \hline
 5.&~$\frac{1}{2}$&~$\frac{1}{2}$ & -$\frac{1}{2}$& -$\frac{1}{2}$&$A_{eem}$ ($\phi_1\Leftrightarrow\chi_1,\phi_2\Leftrightarrow\chi_2$)\\
\hline
6.&~1& ~1  & ~0&~0&$C_{eem}$ ($\phi_1\Leftrightarrow\chi_1,\phi_2\Leftrightarrow\chi_2$)\\
\hline
\hline
 7.&~$\frac{1}{2}$&-$\frac{1}{2}$ & -$\frac{1}{2}$&~$\frac{1}{2}$&$B_{eem}$ ($\phi_1\Leftrightarrow\chi_1,\phi_2\Leftrightarrow\chi_2$)\\
 \hline
8.&~1& ~0 & ~0&~1&$F_{eem}$ ($\phi_1\Leftrightarrow\chi_1,\phi_2\Leftrightarrow\chi_2$)\\
 \hline
\end{tabular}
\caption{Scaling of spinors in EEM sector : [1,2] satisfy $\b_{1,2}-\a_{1,2}=1$;~[3,4] satisfy $\b_{1}-\a_{1}=1$ and $\a_{2}-\b_{2}=1$;~[5,6] satisfy $\a_{1,2}-\b_{1,2}=1$;~[7,8] satisfy $\a_{1}-\b_{1}=1$ and $\b_{2}-\a_{2}=1$ along with (\ref{condeem}).}
\end {table}
The EOM in EEM sector are given as:
\\
$\bullet{~\underline{A_{eem} ~case}}:$
\bea{} 1) &&\p_t\p^j A_{t}^{(1,2)}+\p_{i}(\p^i A^{j(1,2)}-\p^j A^{i(1,2)})=0,~\p_i\p^i A_{t}^{(1,2)}=0,~ i\sigma^{i}\p_{i}\phi_{1,2}=0,\non\\2)&&
\p_i (\p^i A^{j3}-\p^j A^{i3})=0,~ \p_i (\p^i A_{t}^{3}-\p_t A^{i3})-g(\phi^{\dagger}_1 T^{3}_{11}\phi_{1}+\phi^{\dagger}_2 T^{3}_{22}\phi_{2})=0,\non\\3) && i\p_t\phi_{1} +gT^{1}_{12}A^{1}_{t}\phi_2 +gT^{2}_{12}A^{2}_{t}\phi_2 +i\sigma^{i}\p_{i}\chi_{1}=0,\non\\4)&& i\p_t\phi_{2} +gT^{1}_{21}A^{1}_{t}\phi_1 +gT^{2}_{21}A^{2}_{t}\phi_1 +i\sigma^{i}\p_{i}\chi_{2}=0. \non
\eea
$\bullet{~\underline{B_{eem} ~case}}:$
\bea{}1) &&\p_t \p^j A_{t}^{(1,2)}+\p_{i}(\p^i A^{j(1,2)}-\p^j A^{i(1,2)})+g(\phi^{\dagger}_1 \sigma^{j} T^{(1,2)}_{12}\chi_{2}+\chi^{\dagger}_2 \sigma^{j} T^{(1,2)}_{21}\phi_{1})=0,\non\\2)&&
\p_i (\p^i A_{t}^{3}-\p_t A^{i3})-g(\phi^{\dagger}_1 T^{3}_{11}\phi_{1}+\chi^{\dagger}_2 T^{3}_{22}\chi_{2})=0,~\p_i\p^i A_{t}^{(1,2)}=0,\non\\3)&&\p_{i}(\p^i A^{j(3)}-\p^j A^{i(3)})=0,~ i\sigma^{i}\p_{i}\chi_{2}=0,~i\sigma^{i}\p_{i}\phi_{1}=0,\non\\4)&& i\p_t\phi_{1} +i\sigma^{i}\p_{i}\chi_{1}=0,~ i\p_t\chi_{2}+i\sigma^{i}\p_{i}\phi_{2}=0. \non
\eea
$\bullet{~\underline{C_{eem} ~case}}:$
\bea{} 1) &&\p_t\p^j A_{t}^{(1,2)}+\p_{i}(\p^i A^{j(1,2)}-\p^j A^{i(1,2)})=0,~\p_i\p^i A_{t}^{(1,2)}=0,~ i\sigma^{i}\p_{i}\phi_{1,2}=0,
\non\\2)&&\p_i (\p^i A^{j3}-\p^j A^{i3})=0,~ \p_i (\p^i A_{t}^{3}-\p_t A^{i3})=0,\non\\3) && i\p_t\phi_{1} +gT^{1}_{12}A^{1}_{t}\phi_2 +gT^{2}_{12}A^{2}_{t}\phi_2 +i\sigma^{i}\p_{i}\chi_{1}=0,\non\\4)&& i\p_t\phi_{2} +gT^{1}_{21}A^{1}_{t}\phi_1 +gT^{2}_{21}A^{2}_{t}\phi_1 +i\sigma^{i}\p_{i}\chi_{2}=0. \non
\eea
$\bullet{~\underline{F_{eem} ~(Free)~ case}}:$
\bea{}1) &&\p_t\p^j A_{t}^{(1,2)}+\p_{i}(\p^i A^{j(1,2)}-\p^j A^{i(1,2)})=0,~\p_i\p^i A_{t}^{(1,2)}=0,~i\sigma^{i}\p_{i}\phi_{1}=0,\non\\2)&&\p_{i}(\p^i A^{j3}-\p^j A^{i3})=0,~
\p_i (\p^i A_{t}^{3}-\p_t A^{i3})=0,\non\\3)&& i\sigma^{i}\p_{i}\chi_{2}=0, i\p_t\phi_{1} +i\sigma^{i}\p_{i}\chi_{1}=0,~ i\p_t\chi_{2}+i\sigma^{i}\p_{i}\phi_{2}=0. \non
\eea

\subsubsection*{Gauge Transformation:}
To get the gauge transformation, we have to also take the scaling of the gauge parameter $\th^a$ as 
\be{}\th^{1,2} \rightarrow \epsilon^2 \th^{1,2},~~\th^{3} \rightarrow \epsilon \th^{3}.\ee
Therefore, the gauge transformations of gauge fields reads
\be{eemlgl}\delta A_{t}^{1,2}=0,~\delta A^{1,2}_{i}=\frac{1}{g}\p_i \th^{1,2},~\delta A^{3}_{t}=\frac{1}{g}\p_t \th^{3},~\delta A^{3}_{i}=\frac{1}{g}\p_i \th^{3}.\ee
Similarly, for $(\phi,\chi)$ fields, we have
\bea{spsleem}\non \delta\phi_1 =\epsilon (i\th^3 T^{3}_{11}\phi_1)+\epsilon^{\a_2 -\a_1 +2} (i\th^1 T^{1}_{12}\phi_2 +i\th^2 T^{2}_{12}\phi_2),\cr \delta\phi_2 =\epsilon^{\a_1 -\a_2 +2} (i\th^1 T^{1}_{21}\phi_1 +i\th^2 T^{2}_{21}\phi_1)+\epsilon (i\th^3 T^{3}_{22}\phi_2),\cr
\\
\delta\chi_1 =\epsilon (i\th^3 T^{3}_{11}\chi_1)+\epsilon^{\b_2 -\b_1 +2} (i\th^1 T^{1}_{12}\chi_2 + i\th^2 T^{2}_{12}\chi_2),\cr \delta\chi_2 =\epsilon^{\b_1 -\b_2 +2} (i\th^1 T^{1}_{21}\chi_1 +i\th^2 T^{2}_{21}\chi_1)+\epsilon (i\th^3 T^{3}_{22}\chi_2).
\eea  
For a particular set of values of $(\a_1,\a_2,\b_1,\b_2)$ along with $\epsilon \rightarrow 0$ limit, the EOMs will be invariant under these transformations.
\subsection*{EMM sector}
The scaling of the fields in EMM sector are given as follows:
\bea{}A_{i}^{1} \rightarrow \epsilon A_i^{1} ,~ A_t^{1} \rightarrow A_t^{1},~ A_{i}^{2,3} \rightarrow A_i^{2,3} ,~ A_t^{2,3} \rightarrow \epsilon  A_t^{2,3},\non\\ \phi_1 \rightarrow \epsilon^{\a_1}\phi_1 , \phi_2 \rightarrow \epsilon^{\a_2}\phi_2,~\chi_1 \rightarrow \epsilon^{\b_1}\chi_1,~\chi_2 \rightarrow \epsilon^{\b_2}\chi_2 \eea
with the conditions of scaling of spinors as
\be{condemm}\alpha_{i}+\beta_{j}\geq -1 ,~\sum_{i}\alpha_{i}\geq -1,~\sum_{i}\beta_{i}\geq -1,~\alpha_{i}\geq -\frac{1}{2},~\beta_{i}\geq -\frac{1}{2} ~\text{with}~i,j=1,2.\ee
We define some quantities like $A_{emm},B_{emm},...$ and so on.
\begin{table}[htb]
\centering
\label{tabemml} 
\begin{tabular}{|p{1cm}|p{1cm}|p{1cm}|p{1cm}|p{1cm}|p{5cm}|}
 \hline
 \multicolumn{6}{|c|}{EMM sector} \\
 \hline
 &$\a_1$& $\a_2$ &$\b_1$ & $\b_2$& EOM\\
 \hline
 1.&-$\frac{1}{2}$& -$\frac{1}{2}$ &~$\frac{1}{2}$& ~$\frac{1}{2}$&$A_{emm}$\\
  \hline
  2.&~0& ~0  & ~1&~1&$C_{emm}$\\
 \hline
 \hline
 3.&-$\frac{1}{2}$& ~$\frac{1}{2}$  & ~$\frac{1}{2}$   &-$\frac{1}{2}$&$B_{emm}$\\
 \hline
4.&~0& ~1 & ~1&~0&$D_{emm}$\\
 \hline
 \hline
 5.&~$\frac{1}{2}$&~$\frac{1}{2}$ & -$\frac{1}{2}$& -$\frac{1}{2}$&$A_{emm}$ ($\phi_1\Leftrightarrow\chi_1,\phi_2\Leftrightarrow\chi_2$)\\
\hline
6.&~1& ~1  & ~0&~0&$C_{emm}$ ($\phi_1\Leftrightarrow\chi_1,\phi_2\Leftrightarrow\chi_2$)\\
 \hline
 \hline
 7.&~$\frac{1}{2}$&-$\frac{1}{2}$ & -$\frac{1}{2}$&~$\frac{1}{2}$&$B_{emm}$ ($\phi_1\Leftrightarrow\chi_1,\phi_2\Leftrightarrow\chi_2$)\\
 \hline
8.&~1& ~0 & ~0&~1&$D_{emm}$ ($\phi_1\Leftrightarrow\chi_1,\phi_2\Leftrightarrow\chi_2$)\\
 \hline
\end{tabular}
\caption{Scaling of spinors in EMM sector: [1,2] satisfy $\b_{1,2}-\a_{1,2}=1$;~[3,4] satisfy $\b_{1}-\a_{1}=1$ and $\a_{2}-\b_{2}=1$;~[5,6] satisfy $\a_{1,2}-\b_{1,2}=1$;~[7,8] satisfy $\a_{1}-\b_{1}=1$ and $\b_{2}-\a_{2}=1$ along with (\ref{condemm}).}
\end {table}
\\
The EOM in EMM sector are given as:\\
$\bullet{~\underline{A_{emm} ~case}}:$
\bea{} 1) &&\p_t\p^j A_{t}^{1}+\p_{i}(\p^i A^{j1}-\p^j A^{i1}+gA^{i2}A^{j3}-gA^{i3}A^{j2})\non\\&&~~~~~~~~~~~~~~~~~~~~~~~~~~~+gA^{2}_{i}(\p^i A^{j3}-\p^j A^{i3})-gA^{3}_{i}(\p^i A^{j2}-\p^j A^{i2})=0,\non\\2)&& \p_i (\p^i A_{t}^{2}-\p_t A^{i2}+gA^{i3}A_{t}^{1})+gA^{3}_{i}\p^{i}A_{t}^{1}-g(\phi^{\dagger}_1 T^{2}_{12}\phi_{2}+\phi^{\dagger}_2 T^{2}_{21}\phi_{1})=0,\non\\3)&& \p_i (\p^i A_{t}^{3}-\p_t A^{i3}-gA^{i2}A_{t}^{1})-gA^{2}_{i}\p^{i}A_{t}^{1}-g(\phi^{\dagger}_1 T^{3}_{11}\phi_{1}+\phi^{\dagger}_2 T^{3}_{22}\phi_{2})=0,\non\\4)&&
\p_i (\p^i A^{j(2,3)}-\p^j A^{i(2,3)})=0,~\p_i\p^i A_{t}^{1}=0,~ i\sigma^{i}\p_{i}\phi_{1,2}=0,\non\\5) && i\p_t\phi_{1} +gT^{1}_{12}A^{1}_{t}\phi_2 +i\sigma^{i}\p_{i}\chi_{1}=0,\non\\6)&& i\p_t\phi_{2} +gT^{1}_{21}A^{1}_{t}\phi_1 +i\sigma^{i}\p_{i}\chi_{2}=0.
\non
\eea
$\bullet{~\underline{B_{emm} ~case}}:$
\bea{} 1) &&\p_t\p^j A_{t}^{1}+\p_{i}(\p^i A^{j1}-\p^j A^{i1}+gA^{i2}A^{j3}-gA^{i3}A^{j2})+gA^{2}_{i}(\p^i A^{j3}-\p^j A^{i3})\non\\&&~~~~~~~~~~~~~~~~~~~~~~~~~~~~~~~~~-gA^{3}_{i}(\p^i A^{j2}-\p^j A^{i2})+g(\phi^{\dagger}_1 \sigma^{j}T^{1}_{12}\chi_{2}+\chi^{\dagger}_2 \sigma^{j}T^{1}_{21}\phi_{1})=0,\non\\2)&&
\p_i (\p^i A^{j(2,3)}-\p^j A^{i(2,3)})=0,~\p_i\p^i A_{t}^{1}=0,~ i\sigma^{i}\p_{i}\phi_{1}=0,~i\sigma^{i}\p_{i}\chi_{2}=0,\non\\3)&& \p_i (\p^i A_{t}^{3}-\p_t A^{i3}-gA^{i2}A_{t}^{1})-gA^{2}_{i}\p^{i}A_{t}^{1}-g(\phi^{\dagger}_1 T^{3}_{11}\phi_{1}+\chi^{\dagger}_2 T^{3}_{22}\chi_{2})=0,\non\\4)&& \p_i (\p^i A_{t}^{2}-\p_t A^{i2}+gA^{i3}A_{t}^{1})+gA^{3}_{i}\p^{i}A_{t}^{1}=0,\non\\5) && i\p_t\phi_{1} +g\sigma^{i}T^{2}_{12}A^{2}_{i}\chi_2 +i\sigma^{i}\p_{i}\chi_{1}=0,\non\\6)&& i\p_t\chi_{2} +g\sigma^{i}T^{2}_{21}A^{2}_{i}\phi_1 +i\sigma^{i}\p_{i}\phi_{2}=0.
\non
\eea
$\bullet{~\underline{C_{emm} ~case}}:$
\bea{} 1) &&\p_t\p^j A_{t}^{1}+\p_{i}(\p^i A^{j1}-\p^j A^{i1}+gA^{i2}A^{j3}-gA^{i3}A^{j2})\non\\&&~~~~~~~~~~~~~~~~~~~~~~~~~~~+gA^{2}_{i}(\p^i A^{j3}-\p^j A^{i3})-gA^{3}_{i}(\p^i A^{j2}-\p^j A^{i2})=0,\non\\2)&&
\p_i (\p^i A^{j(2,3)}-\p^j A^{i(2,3)})=0,~\p_i\p^i A_{t}^{1}=0,~ i\sigma^{i}\p_{i}\phi_{1,2}=0,\non\\3)&& \p_i (\p^i A_{t}^{2}-\p_t A^{i2}+gA^{i3}A_{t}^{1})+gA^{3}_{i}\p^{i}A_{t}^{1}=0,\non\\4)&& \p_i (\p^i A_{t}^{3}-\p_t A^{i3}-gA^{i2}A_{t}^{1})-gA^{2}_{i}\p^{i}A_{t}^{1}=0,\non\\5) && i\p_t\phi_{1} +gT^{1}_{12}A^{1}_{t}\phi_2 +i\sigma^{i}\p_{i}\chi_{1}=0,\non\\6)&& i\p_t\phi_{2} +gT^{1}_{21}A^{1}_{t}\phi_1 +i\sigma^{i}\p_{i}\chi_{2}=0.
\non
\eea
$\bullet{~\underline{D_{emm} ~case}}:$
\bea{} 1) &&\p_t\p^j A_{t}^{1}+\p_{i}(\p^i A^{j1}-\p^j A^{i1}+gA^{i2}A^{j3}-gA^{i3}A^{j2})\non\\&&~~~~~~~~~~~~~~~~~~~~~~~~~~~+gA^{2}_{i}(\p^i A^{j3}-\p^j A^{i3})-gA^{3}_{i}(\p^i A^{j2}-\p^j A^{i2})=0,\non\\2)&&
\p_i (\p^i A^{j(2,3)}-\p^j A^{i(2,3)})=0,~\p_i\p^i A_{t}^{1}=0,~ i\sigma^{i}\p_{i}\phi_{1}=0,~i\sigma^{i}\p_{i}\chi_{2}=0,\non\\3)&& \p_i (\p^i A_{t}^{2}-\p_t A^{i2}+gA^{i3}A_{t}^{1})+gA^{3}_{i}\p^{i}A_{t}^{1}=0,\non\\4)&& \p_i (\p^i A_{t}^{3}-\p_t A^{i3}-gA^{i2}A_{t}^{1})-gA^{2}_{i}\p^{i}A_{t}^{1}=0,\non\\5) && i\p_t\phi_{1} +g\sigma^{i}T^{2}_{12}A^{2}_{i}\chi_2 +i\sigma^{i}\p_{i}\chi_{1}=0,\non\\6)&& i\p_t\chi_{2} +g\sigma^{i}T^{2}_{21}A^{2}_{i}\phi_1 +i\sigma^{i}\p_{i}\phi_{2}=0.\non
\eea
\subsubsection*{Gauge Transformation:}
We  consider the scaling of $\th^a$  to understand the gauge transformation in this limit. It is given by
\be{}\th^{1} \rightarrow \epsilon^2 \th^{1},~~\th^{2,3} \rightarrow \epsilon \th^{2,3}.\ee
The gauge transformations of the gauge fields becomes
\bea{emmlgl}\non\delta A_{t}^{1}=0,~\delta A^{1}_{i}=\frac{1}{g}\p_i \th^{1}+A^{2}_{i}\th^{3}-A^{3}_{i}\th^{2},~\delta A^{2,3}_{i}=\frac{1}{g}\p_i \th^{2,3},\\
\delta A^{2}_{t}=\frac{1}{g}\p_t \th^{2}-A^{1}_{t}\th^{3},~\delta A^{3}_{t}=\frac{1}{g}\p_t \th^{3}+A^{1}_{t}\th^{2}.
\eea
Similarly, for $(\phi,\chi)$ fields, we get
\bea{spslemm}\non \delta\phi_1 =\epsilon (i\th^3 T^{3}_{11}\phi_1)+\epsilon^{\a_2 -\a_1 +2} (i\th^1 T^{1}_{12}\phi_2)+\epsilon^{\a_2 -\a_1 +1}(i\th^2 T^{2}_{12}\phi_2),\cr \delta\phi_2 =\epsilon^{\a_1 -\a_2 +2} (i\th^1 T^{1}_{21}\phi_1)+\epsilon^{\a_1 -\a_2 +1}(i\th^2 T^{2}_{21}\phi_1)+\epsilon (i\th^3 T^{3}_{22}\phi_2),\cr
\\
\delta\chi_1 =\epsilon (i\th^3 T^{3}_{11}\chi_1)+\epsilon^{\b_2 -\b_1 +2} (i\th^1 T^{1}_{12}\chi_2) +\epsilon^{\b_2 -\b_1 +1} (i\th^2 T^{2}_{12}\chi_2),\cr \delta\chi_2 =\epsilon^{\b_1 -\b_2 +2} (i\th^1 T^{1}_{21}\chi_1) +\epsilon^{\b_1 -\b_2 +1}(i\th^2 T^{2}_{21}\chi_1)+\epsilon (i\th^3 T^{3}_{22}\chi_2).
\eea  
Therefore, the EOMs come out to be invariant under these transformations.

\subsection*{MMM sector}
The scaling of the fields in MMM sector are given as
\bea{}A_{i}^a\rightarrow  A_i^a ,~ A_t^a \rightarrow \epsilon A_t^a,~\phi_1 \rightarrow \epsilon^{\a_1}\phi_1, \phi_2 \rightarrow \epsilon^{\a_2}\phi_2,~\chi_1 \rightarrow \epsilon^{\b_1}\chi_1,~\chi_2 \rightarrow \epsilon^{\b_2}\chi_2. \eea
The conditions for the scaling of spinors in this limit are
\be{condmmm}\alpha_{i}+\beta_{j}\geq -2 ,~\sum_{i}\alpha_{i}\geq -1,~\sum_{i}\beta_{i}\geq -1,~\alpha_{i}\geq -\frac{1}{2},~\beta_{i}\geq -\frac{1}{2} ~\text{with}~i,j=1,2.\ee
Following the above analysis, we define certain quantities like $A_m,B_m,...$ that are related to different sets of EOM.
\begin{table}[htb]
\centering
\label{tabml} 
\begin{tabular}{ |p{1cm}|p{1cm}|p{1cm}|p{1cm}|p{1cm}|p{5cm}|}
 \hline
 \multicolumn{6}{|c|}{MMM sector} \\
 \hline
 &$\a_1$& $\a_2$ &$\b_1$ & $\b_2$& EOM\\
 \hline
 1.&-$\frac{1}{2}$& -$\frac{1}{2}$ &~$\frac{1}{2}$& ~$\frac{1}{2}$&$F_{1m}$\\
  \hline
  2.&-$\frac{1}{2}$ &~0 & ~$\frac{1}{2}$&~1&$F_{2m}$\\
  \hline
 3.&~0&-$\frac{1}{2}$ &~1 &~$\frac{1}{2}$&$F_{2m}$ ($\phi_1\Leftrightarrow\phi_2,\chi_1\Leftrightarrow\chi_2$)\\
 \hline
4.&~0& ~0  & ~1&~1&$F_{3m}$\\
 \hline
 \hline
 5.&-$\frac{1}{2}$& ~$\frac{1}{2}$  & ~$\frac{1}{2}$   &-$\frac{1}{2}$&$A_{m}$\\
  \hline
6.&~0& ~1 & ~1&~0&$B_{m}$\\
 \hline
 \hline
 7.&~$\frac{1}{2}$&~$\frac{1}{2}$ & -$\frac{1}{2}$& -$\frac{1}{2}$&$F_{1m}$ ($\phi_1\Leftrightarrow\chi_1,\phi_2\Leftrightarrow\chi_2$) \\
 \hline
8.&~$\frac{1}{2}$&~1&-$\frac{1}{2}$&~0&$F_{2m}$ ($\phi_1\Leftrightarrow\chi_1,\phi_2\Leftrightarrow\chi_2$)\\
\hline
9.&~1& ~$\frac{1}{2}$  & ~0&-$\frac{1}{2}$&$F_{2m}$ ($\phi_1 \Leftrightarrow \chi_2, \phi_2 \Leftrightarrow \chi_1$)\\
 \hline
10.&~1& ~1  & ~0&~0&$F_{3m}$ ($\phi_1\Leftrightarrow\chi_1,\phi_2\Leftrightarrow\chi_2$)\\
 \hline
 \hline
 11.&~$\frac{1}{2}$&-$\frac{1}{2}$ & -$\frac{1}{2}$&~$\frac{1}{2}$&$A_{m}$ ($\phi_1\Leftrightarrow\chi_1,\phi_2\Leftrightarrow\chi_2$)\\
 \hline
12.&~1& ~0 & ~0&~1&$B_{m}$ ($\phi_1\Leftrightarrow\chi_1,\phi_2\Leftrightarrow\chi_2$)\\
 \hline
\end{tabular}
\caption {Scaling of spinors in MMM sector: [1-4] satisfy $\b_{1,2}-\a_{1,2}=1$;~[5,6] satisfy $\b_{1}-\a_{1}=1$ and $\a_{2}-\b_{2}=1$;~[7-10] satisfy $\a_{1,2}-\b_{1,2}=1$;~[11,12] satisfy $\a_{1}-\b_{1}=1$ and $\b_{2}-\a_{2}=1$ along with (\ref{condmmm}).}
\end {table}
The various sets of EOM in MMM sector are
\\
$\bullet{~\underline{A_{m} ~case}}:$
\bea{} 1) &&\p_{i}(\p^i A^{ja}-\p^j A^{ia})=0,~
\p^i (\p_i A_{t}^{a}-\p_t A_{i}^{a})-g(\phi^{\dagger}_1 T^{a}_{11}\phi_{1}+\chi^{\dagger}_2 T^{a}_{22}\chi_{2})=0,\non\\2)&&i\sigma^{i}\p_{i}\phi_{1}=0,~ i\sigma^{i}\p_{i}\chi_{2}=0,\non\\3) && i\p_t\phi_{1} +g\sigma^{i}T^{a}_{12}A^{a}_{i}\chi_2 +i\sigma^{i}\p_{i}\chi_{1}=0,\non\\4)&& i\p_t\chi_2 +g\sigma^{i}T^{a}_{21}A^{a}_{i}\phi_1 +i\sigma^{i}\p_{i}\phi_{2}=0. \non
\eea
$\bullet{~\underline{B_{m} ~case}}:$
\bea{}1) &&\p_{i}(\p^i A^{ja}-\p^j A^{ia})=0,~
\p_i (\p^i A_{t}^{a}-\p_t A^{ia})=0,~i\sigma^{i}\p_{i}\phi_{1}=0,~i\sigma^{i}\p_{i}\chi_{2}=0,\non\\2)&& i\p_t\phi_{1} +g\sigma^{i}T^{a}_{12}A^{a}_{i}\chi_2 +i\sigma^{i}\p_{i}\chi_{1}=0,\non\\3)&& i\p_t\chi_{2} +g\sigma^{i}T^{a}_{21}A^{a}_{i}\phi_1 +i\sigma^{i}\p_{i}\phi_{2}=0. \non
\eea
\subsubsection*{{Free MMM sector}:}
$\bullet{~\underline{F_{1m} ~case}}:$
\bea{} 1)&&\p_{i}(\p^i A^{ja}-\p^j A^{ia})=0,~
 \p_i (\p^i A_{t}^{a}-\p_t A^{ia})-g(\phi^{\dagger}_m T^{a}_{mn}\phi_{n})=0,\non\\2)&&i\sigma^{i}\p_{i}\phi_{1,2}=0, i\p_t\phi_{1,2}+i\sigma^{i}\p_{i}\chi_{1,2}=0.\non
\eea
$\bullet{~\underline{F_{2m} ~case}}:$
\bea{}1)&& \p_{i}(\p^i A^{ja}-\p^j A^{ia})=0,~
 \p_i (\p^i A_{t}^{a}-\p_t A^{ia})-g(\phi^{\dagger}_1 T^{a}_{11}\phi_{1})=0,\non\\2)&&i\sigma^{i}\p_{i}\phi_{1,2}=0, i\p_t\phi_{1,2}+i\sigma^{i}\p_{i}\chi_{1,2}=0.\non
\eea
 $\bullet{~\underline{F_{3m} ~case}}:$
\bea{}1)&& \p_{i}(\p^i A^{ja}-\p^j A^{ia})=0,~
 \p_i (\p^i A_{t}^{a}-\p_t A^{ia})=0,~ i\sigma^{i}\p_{i}\phi_{1,2}=0\non\\2)&& i\p_t\phi_{1,2}+i\sigma^{i}\p_{i}\chi_{1,2}=0.\non
\eea
\subsubsection*{Gauge Transformation:}
In MMM sector, the gauge transformation can be found by taking the scaling of potentials and $\th^a$ as 
\be{}\th^a \rightarrow \epsilon \th^a.\ee
The gauge transformations of gauge fields in this sector are given as
\be{mlgl}\delta A^{a}_{t}=\frac{1}{g}\p_t \th^{a},~\delta A^{a}_{i}=\frac{1}{g}\p_i \th^{a}.\ee
Similarly, for $(\phi,\chi)$ fields
\bea{spslm}\non \delta\phi_1 =\epsilon (i\th^a T^{a}_{11}\phi_1)+\epsilon^{\a_2 -\a_1 +1} (i\th^a T^{a}_{12}\phi_2),~ \delta\phi_2 =\epsilon^{\a_1 -\a_2 +1} (i\th^a T^{a}_{21}\phi_1)+\epsilon (i\th^a T^{a}_{22}\phi_2),\\ \delta\chi_1 =\epsilon (i\th^a T^{a}_{11}\chi_1)+\epsilon^{\b_2 -\b_1 +1} (i\th^a T^{a}_{12}\chi_2),~ \delta\chi_2 =\epsilon^{\b_1 -\b_2 +1} (i\th^a T^{a}_{21}\chi_1)+\epsilon (i\th^a T^{a}_{22}\chi_2).\cr
\eea  
The EOMs will come out to be invariant under (\ref{mlgl}) and (\ref{spslm}), if we take a particular set of values of the coefficients given in the table.


\bigskip 

\bigskip


\begin{thebibliography}{999}

\bibitem{Simmons-Duffin:2016gjk} 
  D.~Simmons-Duffin,
  ``TASI Lectures on the Conformal Bootstrap,''
  arXiv:1602.07982 [hep-th].

\bibitem{Bagchi:2016geg} 
 A.~Bagchi, M.~Gary and Zodinmawia,
  ``Bondi-Metzner-Sachs bootstrap,''
  Phys.\ Rev.\ D {\bf 96}, no. 2, 025007 (2017)
  doi:10.1103/PhysRevD.96.025007
  [arXiv:1612.01730 [hep-th]].
  
  \bibitem{Bagchi:2017cpu} 
  A.~Bagchi, M.~Gary and Zodinmawia,
  ``The nuts and bolts of the BMS Bootstrap,''
  Class.\ Quant.\ Grav.\  {\bf 34}, no. 17, 174002 (2017)
  doi:10.1088/1361-6382/aa8003
  [arXiv:1705.05890 [hep-th]].

\bibitem{Song:2017czq} 
  W.~Song and J.~Xu,
  ``Correlation Functions of Warped CFT,''
  arXiv:1706.07621 [hep-th].


\bibitem{Taylor:2015glc} 
  M.~Taylor,
  ``Lifshitz holography,''
  Class.\ Quant.\ Grav.\  {\bf 33}, no. 3, 033001 (2016)
  doi:10.1088/0264-9381/33/3/033001
  [arXiv:1512.03554 [hep-th]].

\bibitem{Hagen:1972pd} 
  C.~R.~Hagen,
  ``Scale and conformal transformations in galilean-covariant field theory,''
  Phys.\ Rev.\ D {\bf 5}, 377 (1972).
  doi:10.1103/PhysRevD.5.377

\bibitem{Niederer:1972zz} 
  U.~Niederer,
  ``The maximal kinematical invariance group of the free Schrodinger equation.,''
  Helv.\ Phys.\ Acta {\bf 45}, 802 (1972).
  doi:10.5169/seals-114417

\bibitem{Henkel:1993sg} 
  M.~Henkel,
  ``Schrodinger invariance in strongly anisotropic critical systems,''
  J.\ Statist.\ Phys.\  {\bf 75}, 1023 (1994)
  doi:10.1007/BF02186756
  [hep-th/9310081].

 
    
\bibitem{Bagchi:2009my}
  A.~Bagchi and R.~Gopakumar,
  ``Galilean Conformal Algebras and AdS/CFT,''
  JHEP {\bf 0907}, 037 (2009)
  [arXiv:0902.1385 [hep-th]].
  
 \bibitem{Bagchi:2014ysa} 
  A.~Bagchi, R.~Basu and A.~Mehra,
  ``Galilean Conformal Electrodynamics,''
  JHEP {\bf 1411}, 061 (2014)
  doi:10.1007/JHEP11(2014)061
  [arXiv:1408.0810 [hep-th]].
    
 \bibitem{Bagchi:2015qcw} 
  A.~Bagchi, R.~Basu, A.~Kakkar and A.~Mehra,
  ``Galilean Yang-Mills Theory,''
  JHEP {\bf 1604}, 051 (2016)
  doi:10.1007/JHEP04(2016)051
  [arXiv:1512.08375 [hep-th]].
  
\bibitem{LBLL}
 M. Le Bellac and J.-M. L\'evy-Leblond,
``Galilean Electromagnetism,''
 Nuovo Cimento {\bf 14B} (1973), 217    

   
\bibitem{Beisert:2010jr} 
  N.~Beisert, C.~Ahn, L.~F.~Alday, Z.~Bajnok, J.~M.~Drummond, L.~Freyhult, N.~Gromov and R.~A.~Janik {\it et al.},
  ``Review of AdS/CFT Integrability: An Overview,''
  Lett.\ Math.\ Phys.\  {\bf 99}, 3 (2012)
  [arXiv:1012.3982 [hep-th]].

\bibitem{Brambilla:2004jw} 
  N.~Brambilla, A.~Pineda, J.~Soto and A.~Vairo,
  ``Effective field theories for heavy quarkonium,''
  Rev.\ Mod.\ Phys.\  {\bf 77}, 1423 (2005)
  doi:10.1103/RevModPhys.77.1423
  [hep-ph/0410047].
  
 \bibitem{Soto:2006zs} 
  J.~Soto,
  ``Overview of non-relativistic QCD,''
  Eur.\ Phys.\ J.\ A {\bf 31}, 705 (2007)
  doi:10.1140/epja/i2006-10255-9
  [nucl-th/0611055].
  
 \bibitem{Duval:2014uoa} 
  C.~Duval, G.~W.~Gibbons, P.~A.~Horvathy and P.~M.~Zhang,
  ``Carroll versus Newton and Galilei: two dual non-Einsteinian concepts of time,''
  Class.\ Quant.\ Grav.\  {\bf 31}, 085016 (2014)
  [arXiv:1402.0657 [gr-qc]].
 
  

\bibitem{Festuccia:2016caf} 
  G.~Festuccia, D.~Hansen, J.~Hartong and N.~A.~Obers,
  ``Symmetries and Couplings of non-relativistic Electrodynamics,''
  JHEP {\bf 1611}, 037 (2016)
  doi:10.1007/JHEP11(2016)037
  [arXiv:1607.01753 [hep-th]].
  
 \bibitem{Bergshoeff:2015sic} 
  E.~Bergshoeff, J.~Rosseel and T.~Zojer,
  ``Non-relativistic fields from arbitrary contracting backgrounds,''
  Class.\ Quant.\ Grav.\  {\bf 33}, no. 17, 175010 (2016)
  doi:10.1088/0264-9381/33/17/175010
  [arXiv:1512.06064 [hep-th]].
  
  \bibitem{Bleeken:2015ykr} 
  D.~V.~d.~Bleeken and C.~Yunus,
  ``Newton-Cartan, Galileo-Maxwell and Kaluza-Klein,''
  arXiv:1512.03799 [hep-th].

  



\bibitem{Jackiw:2011vz} 
  R.~Jackiw and S.~-Y.~Pi,
  ``Tutorial on Scale and Conformal Symmetries in Diverse Dimensions,''
  J.\ Phys.\ A {\bf 44}, 223001 (2011)
  [arXiv:1101.4886 [math-ph]].
  



    \bibitem{Bagchi:2009ca}
  A.~Bagchi and I.~Mandal,
  ``On Representations and Correlation Functions of Galilean Conformal
  Algebras,''
  Phys.\ Lett.\  B {\bf 675}, 393 (2009)
  [arXiv:0903.4524 [hep-th]].

 
 
\bibitem{Bagchi:2009pe} 
  A.~Bagchi, R.~Gopakumar, I.~Mandal and A.~Miwa,
  ``GCA in 2d,''
  JHEP {\bf 1008}, 004 (2010)
  [arXiv:0912.1090 [hep-th]].
  
  
 


\bibitem{Duval:2014uva} 
  C.~Duval, G.~W.~Gibbons and P.~A.~Horvathy,
  ``Conformal Carroll groups and BMS symmetry,''
  Class.\ Quant.\ Grav.\  {\bf 31}, 092001 (2014)
  [arXiv:1402.5894 [gr-qc]].
    
  \bibitem{Duval:2009vt} 
  C.~Duval and P.~A.~Horvathy,
  ``non-relativistic conformal symmetries and Newton-Cartan structures,''
  J.\ Phys.\ A {\bf 42}, 465206 (2009)
  [arXiv:0904.0531 [math-ph]].
  
\bibitem{deAzcarraga:2009ch} 
  J.~A.~de Azcarraga and J.~Lukierski,
  ``Galilean Superconformal Symmetries,''
  Phys.\ Lett.\ B {\bf 678}, 411 (2009)
  [arXiv:0905.0141 [math-ph]].

   \bibitem{Sakaguchi:2009de} 
  M.~Sakaguchi,
  ``Super Galilean conformal algebra in AdS/CFT,''
  J.\ Math.\ Phys.\  {\bf 51}, 042301 (2010)
  [arXiv:0905.0188 [hep-th]].

 \bibitem{Bagchi:2009ke} 
  A.~Bagchi and I.~Mandal,
  ``Supersymmetric Extension of Galilean Conformal Algebras,''
  Phys.\ Rev.\ D {\bf 80}, 086011 (2009)
  [arXiv:0905.0580 [hep-th]].
  

\bibitem{Jensen:2014hqa} 
  K.~Jensen,
  ``Anomalies for Galilean fields,''
  arXiv:1412.7750 [hep-th].

\bibitem{Jain:2015jla} 
  A.~Jain,
  ``Galilean Anomalies and Their Effect on Hydrodynamics,''
  arXiv:1509.05777 [hep-th].
   

 

  
 \bibitem{Bondi:1962px} 
  H.~Bondi, M.~G.~J.~van der Burg and A.~W.~K.~Metzner,
  ``Gravitational waves in general relativity. 7. Waves from axisymmetric isolated systems,''
  Proc.\ Roy.\ Soc.\ Lond.\ A {\bf 269}, 21 (1962).
  
  \bibitem{Sachs:1962zza} 
  R.~Sachs,
  ``Asymptotic symmetries in gravitational theory,''
  Phys.\ Rev.\  {\bf 128}, 2851 (1962).
  
  \bibitem{Strominger:2017zoo} 
  A.~Strominger,
  ``Lectures on the Infrared Structure of Gravity and Gauge Theory,''
  arXiv:1703.05448 [hep-th].

 \bibitem{Barnich:2006av} 
  G.~Barnich and G.~Compere,
  ``Classical central extension for asymptotic symmetries at null infinity in three spacetime dimensions,''
  Class.\ Quant.\ Grav.\  {\bf 24}, F15 (2007)
  [gr-qc/0610130].

\bibitem{Bagchi:2010eg} 
  A.~Bagchi,
  ``The BMS/GCA correspondence,''
  Phys.\ Rev.\ Lett.\  {\bf 105}, 171601 (2010)
  [arXiv:1006.3354 [hep-th]].

\bibitem{Bagchi:2012yk} 
  A.~Bagchi, S.~Detournay and D.~Grumiller,
  ``Flat-Space Chiral Gravity,''
  Phys.\ Rev.\ Lett.\  {\bf 109}, 151301 (2012)
  [arXiv:1208.1658 [hep-th]].

\bibitem{Bagchi:2012cy} 
  A.~Bagchi and R.~Fareghbal,
  ``BMS/GCA Redux: Towards Flatspace Holography from non-relativistic Symmetries,''
  JHEP {\bf 1210}, 092 (2012)
  doi:10.1007/JHEP10(2012)092
  [arXiv:1203.5795 [hep-th]].
  


\bibitem{Bagchi:2012xr} 
  A.~Bagchi, S.~Detournay, R.~Fareghbal and J.~Simon,
  ``Holography of 3d Flat Cosmological Horizons,''
  Phys.\ Rev.\ Lett.\  {\bf 110}, 141302 (2013)
  [arXiv:1208.4372 [hep-th]].

\bibitem{Barnich:2012xq} 
  G.~Barnich,
  ``Entropy of three-dimensional asymptotically flat cosmological solutions,''
  JHEP {\bf 1210}, 095 (2012)
  [arXiv:1208.4371 [hep-th]].

\bibitem{Bagchi:2014iea} 
  A.~Bagchi, R.~Basu, D.~Grumiller and M.~Riegler,
  ``Entanglement entropy in Galilean conformal field theories and flat holography,''
  Phys.\ Rev.\ Lett.\  {\bf 114}, no. 11, 111602 (2015)
  doi:10.1103/PhysRevLett.114.111602
  [arXiv:1410.4089 [hep-th]].


\bibitem{Barnich:2012rz} 
  G.~Barnich, A.~Gomberoff and H.~A.~Gonzalez,
  ``BMS$_3$ invariant two dimensional field theories as flat limit of Liouville,''
  arXiv:1210.0731 [hep-th].

\bibitem{Bagchi:2016bcd} 
  A.~Bagchi, R.~Basu, A.~Kakkar and A.~Mehra,
  ``Flat Holography: Aspects of the dual field theory,''
  JHEP {\bf 1612}, 147 (2016)
  doi:10.1007/JHEP12(2016)147
  [arXiv:1609.06203 [hep-th]].

\bibitem{Riegler:2016hah} 
  M.~Riegler,
  ``How General Is Holography?,''
  arXiv:1609.02733 [hep-th].
  
  
\bibitem{Bagchi:2013bga} 
  A.~Bagchi,
  ``Tensionless Strings and Galilean Conformal Algebra,''
  JHEP {\bf 1305}, 141 (2013)
  [arXiv:1303.0291 [hep-th]].
  
\bibitem{Bagchi:2015nca} 
  A.~Bagchi, S.~Chakrabortty and P.~Parekh,
  ``Tensionless Strings from Worldsheet Symmetries,''
  JHEP {\bf 1601}, 158 (2016)
  doi:10.1007/JHEP01(2016)158
  [arXiv:1507.04361 [hep-th]].
  
 \bibitem{Casali:2016atr} 
  E.~Casali and P.~Tourkine,
  ``On the null origin of the ambitwistor string,''
  JHEP {\bf 1611}, 036 (2016)
  doi:10.1007/JHEP11(2016)036
  [arXiv:1606.05636 [hep-th]].
  

\end{thebibliography}
\end{document}